\documentclass[10pt,twocolumn,letterpaper]{article}

\usepackage[pagenumbers]{cvpr} % To force page numbers, e.g. for an arXiv version

\definecolor{cvprblue}{rgb}{0.21,0.49,0.74}
\usepackage[pagebackref,breaklinks,colorlinks,allcolors=cvprblue]{hyperref}

\usepackage{multirow}
\usepackage{tikz}
\usepackage[table]{xcolor}
\definecolor{mygray}{gray}{0.85}
\definecolor{babyblue}{rgb}{0.80, 0.90, 0.97}

\definecolor{bluek1}{rgb}{0.95, 0.98, 1.00}
\definecolor{bluek4}{rgb}{0.90, 0.96, 0.99}
\definecolor{bluek8}{rgb}{0.85, 0.94, 0.98}
\usepackage{threeparttable}
\usepackage{colortbl}
\usepackage{wrapfig}
\usepackage{comment}

\usepackage{pifont}
\newcommand{\xmark}{\ding{55}}
\usepackage{cuted}
\usepackage{bbding}
\usepackage{dsfont}

\usepackage{fontawesome5}

\usepackage[accsupp]{axessibility}  % Improves PDF readability for those with disabilities.

\newcommand{\methodname}{FLAC}
\newcommand{\AGname}{AGREE}

\newcommand{\unc}[1]{\makebox[0pt][l]{\scriptsize$_{\pm #1}$}}
\newcommand{\abl}[0]{$^{\text{\tiny\faCut}}$}
\newcommand{\amandine}[1]{#1}
\newcommand{\amandinec}[1]{}

\usepackage{placeins}

\def\paperID{}
\def\confName{CVPR}
\def\confYear{2026}

\title{Few-shot Acoustic Synthesis with Multimodal Flow Matching}

\author{Amandine Brunetto \\
Center for Robotics, Mines Paris - PSL University\\
{\tt\small \url{https://amandinebtto.github.io/}}
}

\begin{document}
\newcolumntype{P}[1]{>{\hspace{0pt}}c<{\hspace{#1}}}
\maketitle

% For big figure at the begining fo the document
\begin{strip}
    \centering    
    \vspace{-45pt}
    
    \includegraphics[width=\linewidth]{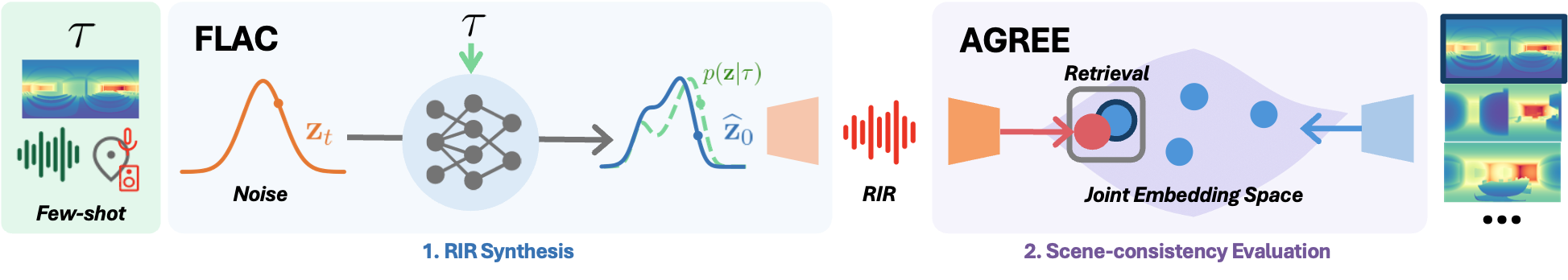}
    \vspace{-20pt}

    \captionof{figure}{\textbf{Few-shot \underline{fl}ow-matching \underline{ac}oustic synthesis (FLAC) and scene-consistency evaluation:} 
    Given a few-shot multimodal context $\tau$, including a depth map, an acoustic observation, and sensor poses, \methodname~uses a diffusion transformer trained with flow matching to generate room impulse responses (RIRs) in novel rooms.
    Unlike prior deterministic approaches, \methodname~models the distribution of plausible RIRs under sparse scene context, capturing acoustic uncertainty. Even with one shot, \methodname~outperforms 8-shot state-of-the-art methods.
    To assess generation quality, we introduce AGREE, a CLIP-style audio-geometry embedding that aligns both modalities in a shared latent space, enabling scene-consistency evaluation through retrieval and distributional metrics.
    }
    \label{fig: FS_concept}
\end{strip}

\begin{abstract}
Generating audio that is acoustically consistent with a scene is essential for immersive virtual environments. 
Recent neural acoustic field methods enable spatially continuous sound rendering but remain scene-specific, requiring dense audio measurements and costly training for each environment. 
Few-shot approaches improve scalability across rooms but still rely on multiple recordings and, being deterministic, fail to capture the inherent uncertainty of scene acoustics under sparse context.
We introduce ~\underline{fl}ow-matching \underline{ac}oustic generation (FLAC), a probabilistic method for few-shot acoustic synthesis that models the distribution of plausible room impulse responses (RIRs) given minimal scene context. FLAC leverages a diffusion transformer trained with a flow-matching objective to generate RIRs at arbitrary positions in novel scenes, conditioned on spatial, geometric, and acoustic cues.
FLAC outperforms state-of-the-art eight-shot baselines with one-shot on both the AcousticRooms and Hearing Anything Anywhere datasets. 
To complement standard perceptual metrics, we further introduce AGREE, a joint \underline{a}coustic–\underline{g}eomet\underline{r}y \underline{e}mb\underline{e}dding, enabling geometry-consistent evaluation of generated RIRs through retrieval and distributional metrics.
This work is the first to apply generative flow matching to explicit RIR synthesis, establishing a new direction for robust and data-efficient acoustic synthesis. Project page: \url{https://amandinebtto.github.io/FLAC/}

\end{abstract}

\section{Introduction}
\label{sec:intro}

Every room shapes the way we hear: a lecture hall amplifies a speaker’s voice, while a cathedral envelops sound in lingering reverberation. Reproducing these rich auditory experiences is essential for creating virtual, immersive environments, where users expect sound to reflect the space.

The acoustic properties of a room are encapsulated by Room Impulse Responses (RIRs), which describe the sound propagation between source-receiver pairs. RIRs allows for auralization, \ie, transferring a room's acoustic signature onto any sound. However, accurately modeling RIRs is challenging because they depend on complex interactions between geometry, materials, and source-listener positions. 

Recently, neural acoustic fields \cite{NAF, INRAS, AVNeRF, NeRAF, AVGS, AVCloud, HAA} have enabled spatially continuous RIRs rendering in a scene. However, they must be trained for each environment using extensive RIR recordings. More scalable solutions require models that can generate RIRs in novel rooms, with minimal data and without retraining.

A handful of works have explored few-shot acoustic synthesis \cite{fewshot, MAGIC, liu2025haae}. These methods generate RIRs in novel environments using only a sparse set of information (\eg, depth maps, RGB images, sensor poses, and 8 to 20 RIR recordings) without scene-specific retraining.
With limited knowledge about a new scene's characteristics, there is no single, deterministic possible RIR: few-shot generalization is an inherently ambiguous problem. Yet, existing few-shot methods overlook this uncertainty, producing only a unique deterministic prediction.

To address this, we propose \methodname, a conditional generative model for few-shot acoustic synthesis based on flow matching \cite{lipman2023flow}. This framework extends diffusion models \cite{ho2020denoising, song2020score} with increased performance and versatility and has demonstrated strong performance in audio \cite{liu2024audioldm2, hung2024tangoflux, etta} and images \cite{esser2024scaling} generation.  
\methodname~is, to the best of our knowledge, the first application of generative flow matching to explicit RIR synthesis. Rather than learning a deterministic mapping, our model estimates a distribution of plausible RIRs given sparse scene context, explicitly capturing the uncertainty inherent in few-shot scenarios. We condition the generation on multimodal context, including scene geometry around the receiver, sensor poses, and a minimal set of RIR recordings. 
By formulating few-shot acoustic synthesis as a conditional generative task, we enable scene-consistent sound generation in novel environments even from only one audio measurement.

To assess generation quality, we complement traditionally used perceptual metrics by introducing a set of scene consistency metrics that ensure the predicted RIR matches the scene's geometry. 
To this end, we introduce AGREE (\underline{A}coustic-\underline{G}eomet\underline{R}y \underline{E}mb\underline{E}dding), a CLIP-style \cite{CLIP} dual-encoder network that aligns RIRs and scene geometry in a shared latent space. This alignment enables zero-shot audio and geometry retrieval. We leverage this shared space to provide a geometry-consistent evaluation framework through both retrieval-based scores and distributional metrics. 

We evaluate \methodname~on the large-scale synthetic AcousticRooms \cite{liu2025haae} dataset. It achieves state-of-the-art RIR synthesis performance, demonstrating generalization across novel source-receiver positions within known rooms, as well as in entirely unseen environments. 
We also validate our model's real-world capabilities through sim-to-real transfer on the Hearing-Anything-Anywhere \cite{HAA} dataset. 
On both datasets, \methodname~outperforms current state-of-the-art methods based on 8 audio recordings with a single one.

\noindent In summary, our main contributions are as follows:
\begin{itemize}
    \item We propose \methodname~the first conditional generative model for few-shot RIR synthesis based on flow matching.
    This approach accounts for the inherent uncertainty of acoustics given sparse scene context, leading to more robust predictions. 
    
    \item Our approach sets a new state-of-the-art on the AcousticRooms and Hearing-Anything-Anywhere datasets, generalizing to both novel source-listener pairs and environments. \methodname~outperforms prior work with $8\times$ fewer RIR recordings.
    
    \item We introduce AGREE, a joint acoustic-geometry embedding space, and propose new scene-consistency metrics that evaluate how well predicted RIRs align with the scene geometry.
    
\end{itemize}

\section{Related Work}

\paragraph{Audio-visual learning.}
Audio-visual learning enhances both acoustic and vision-related tasks, including audio spatialization \citep{2.5D, zhou2020sep, garg2023visually, morgado2018self, vasudevan2020semantic, kimvisage}, de-reverberation \citep{chen2023learning, DereverbAdVerb}, RIR prediction \citep{image2reverb, chen2022visual, AcMatching, fewshot, AVNeRF, NACF, liu2025haae}, depth estimation \citep{BV, BVDataset, BITD, zhu2022beyond, echodiffusion}, navigation \citep{younes2023catch, visualechoes, chen2023sound, AVNav, gan2020look, gao2023sonicverse}, floorplan reconstruction \citep{chat2map, floorplan}, and pose estimation \citep{chen2023sound}. 
\methodname~extends this line of work by leveraging depth information for scene-consistent RIR generation.

\vspace{-5pt}

\paragraph{Neural acoustic fields.}
Neural acoustic fields render RIRs at novel poses by implicitly learning a mapping from spatial coordinates to the room’s acoustic field. Some approaches incorporate physical acoustic models \cite{INRAS, HAA}, others infer local geometry \cite{NAF}, exploit vision cues \cite{AVNeRF, NACF, RAF, AVCloud} or use NeRF \cite{mildenhall2021nerf} and Gaussian splatting-based \cite{kerbl20233d} representations \cite{NeRAF, AVGS}. However, these methods remain scene-specific, requiring dense recordings and retraining for each new environment.

\vspace{-5pt}

\paragraph{Few-shot acoustic synthesis.}
Few-shot methods generalize across scenes using sparse observations. FewShotRIR \citep{fewshot} uses 20 RGB, depth and binaural audio inputs. 
MAGIC \citep{MAGIC} adds semantics by extracting features with a segmentation-pretrained U-Net \cite{unet}. 
More recently, xRIR \citep{liu2025haae} reduces inputs to eight audio recording and a panoramic depth map, and introduces the AcousticRooms dataset specifically designed for cross-room synthesis.
All prior methods treat few-shot RIR prediction as a deterministic mapping, overlooking the ambiguity of the task. 
By using generative flow matching, \methodname~captures the distribution of plausible RIRs given sparse context, improving generalization to new scenes even with one-shot.  

\vspace{-5pt}

\paragraph{Audio diffusion and flow matching.}
Diffusion-based models have advanced text-to-audio generation across speech, music, and general sound \cite{AudioLDM, liu2024audioldm2, huang2023makeanaudio, ghosal2023tango, majumder2024tango2, stableaudio, stableaudioopen}.  
Flow matching further improves synthesis efficiency \cite{hung2024tangoflux, etta, guo2024voiceflow}. \cite{liangbinauralflow} recently achieved speech binauralization via flow matching.
Building on these advances, we adapt generative flow matching to RIR synthesis conditioned on few-shot scene context. 

\vspace{-5pt}

\paragraph{Joint embedding models across modalities.}
Joint embedding models align data from different modalities in a shared representation space.
CLIP \cite{CLIP} pioneered this for image-text, later extended to audio-visual \cite{AVID, MAAVT, audioCLIP, avrir}, audio-text \cite{CLAP, audioCLIP}, and audio with diverse sensory modalities \cite{imagebind}, enabling zero-shot cross-modal retrieval. 
Standard audio embeddings cannot be applied directly to RIRs, which differ substantially. 
We introduce AGREE, a joint embedding space for RIRs and scene geometry, allowing acoustic-geometry consistency evaluation.

\begin{figure*}[t]
    \centerline{\includegraphics[width=\linewidth]{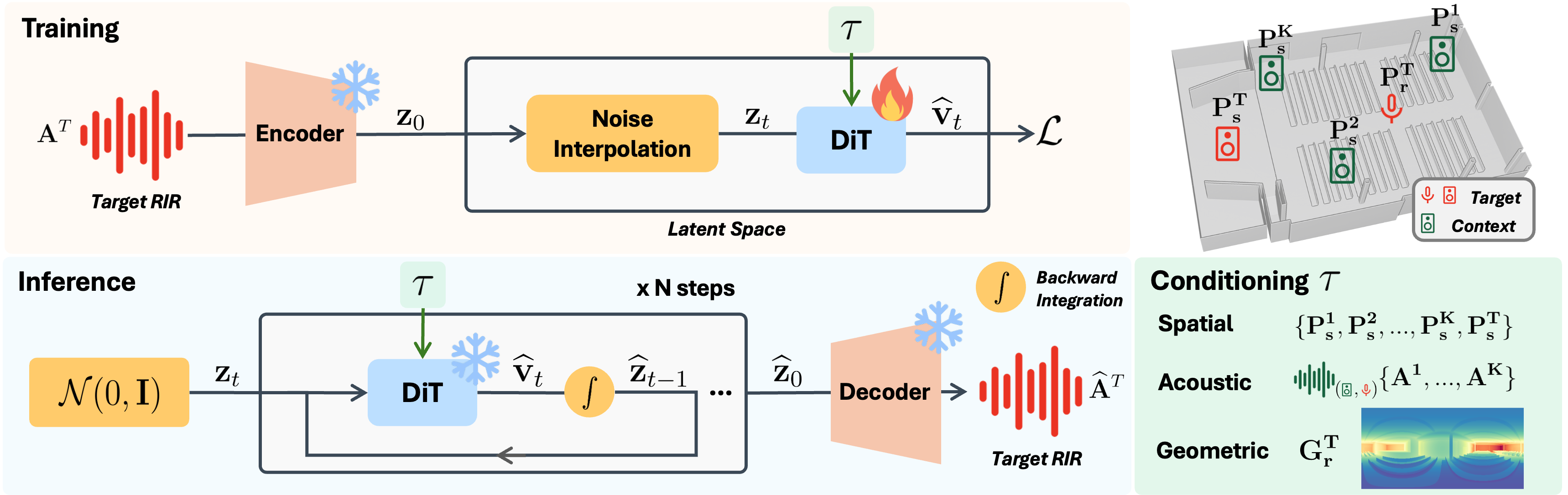}}
    \caption{\textbf{Training and inference pipelines of \methodname:} During training, a pre-trained VAE encodes ground-truth RIRs into latents $\mathbf{z}_0$. Latents are linearly interpolated with noise to form  $\mathbf{z}_t$. A DiT is trained to predict the velocity $\widehat{\mathbf{v}}_t$ that transports $\mathbf{z}_t$ toward the original data distribution. 
    At inference, RIRs are generated from random noise, guided by the few-shot spatial, geometric and acoustic context. 
    }
    \label{fig: FS_Overview}
    \vspace{-5pt}
\end{figure*}

\section{Method}
\methodname~is a conditional latent generative model \cite{rombach2022high} trained with flow matching \cite{lipman2023flow, albergo2023building} (\cref{subsec: LatentRF}) to synthesize RIRs from few-shot scene information. It comprises: (i) a variational autoencoder (\cref{subsec: VAE}), (ii) a multimodal conditioner (\cref{subsec: MMCond}), and (iii) a diffusion transformer (\cref{subsec: DiT}). \cref{fig: FS_Overview} provides an overview of the method.

\subsection{Latent Flow Matching} \label{subsec: LatentRF}

\paragraph{Ambiguity in few-shot synthesis.}
Estimating RIRs across diverse environments and sensor poses is challenging, as they depend on many intertwined factors. With limited scene information, multiple RIRs can be equally plausible for the same source-receiver configuration. For instance, even with precise geometry knowledge, missing material properties introduces ambiguity: whether the floor is carpeted or wooden alters the acoustics. 

We address the inherently ambiguous problem of few-shot RIR synthesis: Our goal is to predict monaural, omnidirectional RIRs at arbitrary source-receiver pairs in unseen environments, given minimal scene context. By using a stochastic generative model, we aim to capture the uncertainty inherent to RIR prediction under sparse observations.

\paragraph{Training.}
We train \methodname~using the rectified flow matching formulation \cite{liu2023flow, liu2022rectified}, which linearly interpolates data and noise. This approach straightens the transport paths between distributions, reducing the number of integration steps at inference. 

The goal is to capture the relationship between a RIR and its spatial, geometric, and acoustic context. 
To this end, we sample target RIRs with their associated context $(A^T, \boldsymbol{\tau})$ from the dataset. Each RIR is encoded into a latent representation $\mathbf{z}_0$, which is linearly interpolated with Gaussian noise $\boldsymbol{\epsilon} \sim \mathcal{N}(0, \mathbf{I})$ to produce a noisy latent $\mathbf{z}_t$:
\begin{equation}
    \mathbf{z}_t = (1 - t)\, \mathbf{z}_0 + t\, \boldsymbol{\epsilon} ,
\end{equation}
where the timestep $t \in [0, 1]$ controls the noise level.
Timesteps are sampled by drawing 
$\alpha \sim \mathcal{N}(-1.2, 2^2)$ and mapping it to $t$ using a sigmoid:
\begin{equation}
    t = \sigma(-\boldsymbol{\alpha}) = \frac{1}{1 + e^{\boldsymbol{\alpha}}}.
\end{equation}
This schedule emphasizes on moderately noisy latents ($t \approx 0.7{-}0.8$), which we found to improve performance. Comparisons of noise sampling strategies are provided in Appendix \labelcref{subsub: timestepsampler}.

The model $u(\mathbf{z}_t, t, \boldsymbol{\tau})$ is trained to predict the velocity field $\mathbf{v}_t$ 
\begin{equation}
    \mathbf{v}_t = \frac{d\mathbf{z}_t}{dt} = \boldsymbol{\epsilon} - \mathbf{z}_0,
\end{equation}
using the following objective:
\begin{equation}
    \mathcal{L}_{\text{RFM}} = \mathbb{E}_{\mathbf{z}_0, \boldsymbol{\epsilon}, t, \boldsymbol{\tau}} \Big[\, \|\, u(\mathbf{z}_t, t, \boldsymbol{\tau}) - \mathbf{v}_t \|^2 \Big].
\end{equation}

\paragraph{Inference.}
We employ classifier-free guidance \cite{ho2021classifier}, allowing the model to learn both conditional and unconditional distributions by randomly dropping the conditioning during training. 

At inference, the guided velocity prediction is given by
\begin{equation}
    \hat{u}(\mathbf{z}_t, t, \boldsymbol{\tau}) = u(\mathbf{z}_t, t, \varnothing) \, + \, \omega \,\big[u(\mathbf{z}_t, t, \boldsymbol{\tau}) - u(\mathbf{z}_t, t, \varnothing)\big],
\end{equation}
where $\omega > 0$ controls the conditioning strength, and $u(\mathbf{z}_t, t, \varnothing)$ denotes the unconditional prediction.

RIRs are generated by solving the ordinary differential equation (ODE) backward, starting from Gaussian noise $\boldsymbol{\epsilon}$ and integrating the velocity field from $t = 1$ to $t = 0$:
\begin{equation}
\mathbf{z}_{t - dt} = \mathbf{z}_t + \hat{u}(\mathbf{z}_t, t, \boldsymbol{\tau}) \, dt.
\end{equation}

\begin{figure}[t]
\centerline{\includegraphics[width=\columnwidth]{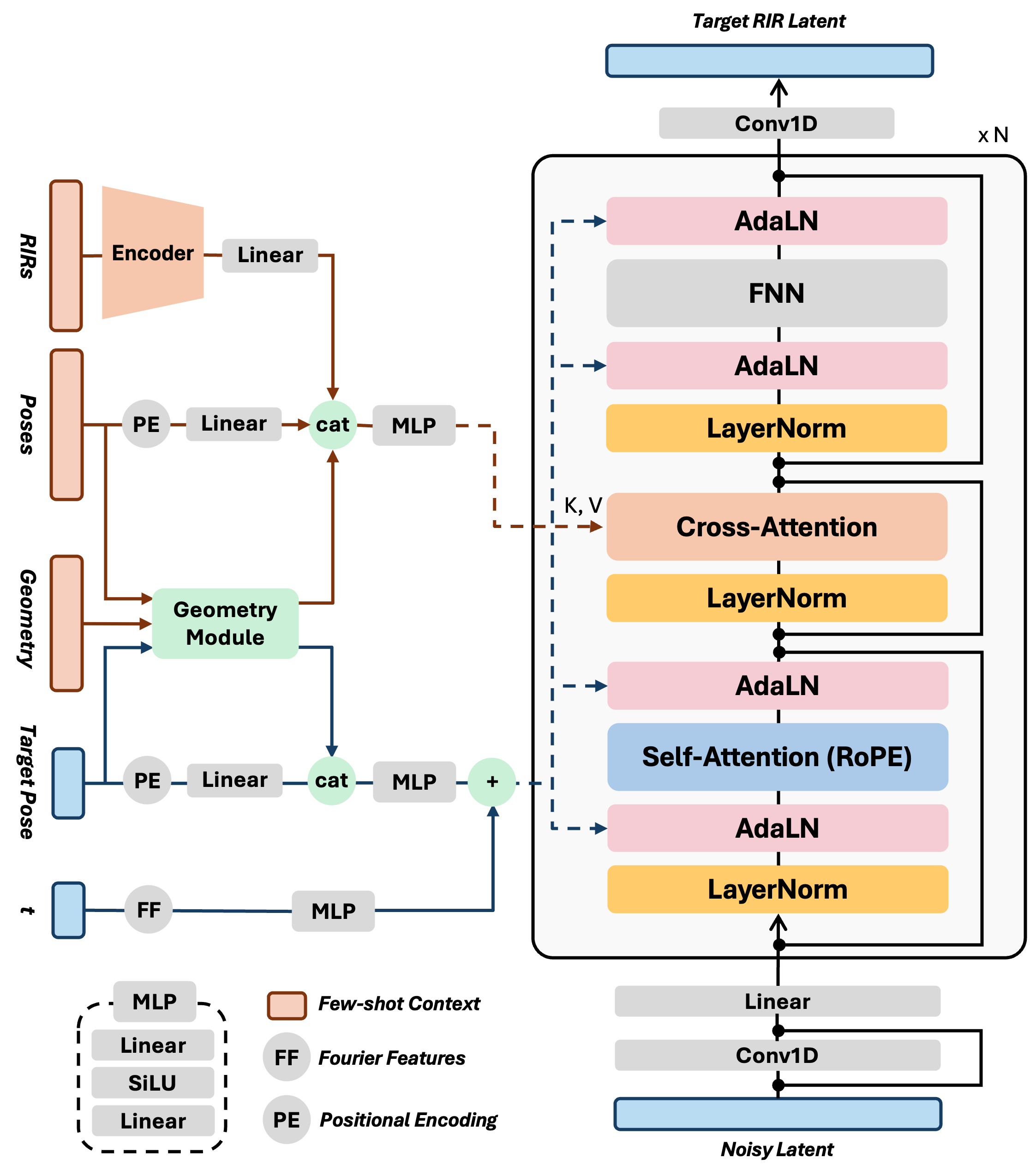}}
    \caption{\textbf{\methodname~diffusion transformer:} The noise timestep $t$ and the target RIR pose are injected via AdaLN. Acoustic, spatial and geometric context are provided through cross-attention.
    }
    \label{fig: FS_DiT}
\end{figure}

\subsection{VAE} \label{subsec: VAE}
We train a variational autoencoder (VAE) to compress RIR waveforms into $\mathbf{z}_0$. 
The encoder consists of four convolutional blocks, each performing downsampling and channel expansion with strided convolutions. Before each downsampling block, we apply ResNet-style layers with dilated convolutions and Snake activations \cite{ziyin2020neural}. The bottleneck has a latent feature dimension of 32, and the decoder mirrors the encoder. All convolutions are weight-normalized \cite{salimans2016weight} and the output passes through a $\tanh$ activation to match the RIR amplitude range.

We found pre-trained audio embeddings unsuitable for latent flow matching.
Obtaining a compact RIR representation is challenging as it must preserve precise temporal and spectral structure.
To achieve this, we train the VAE with complementary objectives: a multiresolution STFT loss $\mathcal{L}_{\text{MR}}$ \cite{steinmetz2020auraloss,yamamoto2020parallel} combining spectral convergence, spectral and energy decay terms; 
an adversarial hinge loss $\mathcal{L}_{\text{adv}}$; a feature-matching loss $\mathcal{L}_{\text{feat}}$ using Encodec \cite{Encodec} multi-scale STFT discriminator; and a KL divergence loss $\mathcal{L}_{\text{KL}}$ to regularize the latent space.
The final objective is:
\begin{equation}
   \mathcal{L} = \mathcal{L}_{\text{MR}} + \mathcal{L}_{\text{adv}} + \mathcal{L}_{\text{feat}} + \mathcal{L}_{\text{KL}} 
\end{equation}
Details on the implementation, individual loss terms, and hyperparameters are provided in Appendix \labelcref{sec: SM_VAE}.

\subsection{Multimodal Conditioning} \label{subsec: MMCond}
\methodname~generates RIRs at a target source-receiver pair $(P_s^T,P_r^T)$ based on multimodal scene context $\boldsymbol{\tau}$:
\begin{itemize}
    \item \textbf{Acoustic:} RIRs measured at the target receiver $P_r^T$ from $K$ different source positions, $\mathbf{A} = \{ A^{1}, \dots, A^{K} \}$, capturing key room acoustic properties.
    \item \textbf{Spatial:} Corresponding source positions $\mathbf{S} = \{ P_{s}^{1}, \dots, P_{s}^{K}\}$ and the target source position $P_s^T$.
    \item \textbf{Geometric:} A panoramic depth map $\mathbf{G}_r^T$ captured at the target receiver pose $P_r^T$, describing local room structure and surfaces.
\end{itemize}
Below, we detail how each modality is processed.

\vspace{-10pt}
\paragraph{Acoustic.}
Similar to \cite{liu2025haae, fewshot} each of the $K$ context RIRs is transformed into a magnitude spectrogram and encoded with a ResNet-18 backbone \cite{ResNet}, trained jointly with the rest of the model. 
The encoder outputs a 512-dimensional embedding per RIR, capturing key acoustic properties. 

\vspace{-10pt}
\paragraph{Spatial.} 
Since the receiver is shared between the context and target RIRs, we express all source poses in the receiver’s local coordinate frame and omit $P_r^T$ (the origin).
The resulting 3D coordinates are encoded with sinusoidal positional embeddings and projected into a high-dimensional feature space through a linear layer.

\vspace{-10pt}
\paragraph{Geometric.} 
We condition on the geometry surrounding the receiver to capture the location and shape of nearby surfaces.
A panoramic depth map captured at $P_r^T$ is converted into an image containing 3D coordinates via equirectangular projection. 
Following \cite{liu2025haae}, we compute reflection maps by subtracting each source position (target and $K$ context) expressed in the receiver's frame from these 3D coordinates. 
DINOv3 \cite{dinov3} Vision Transformer (ViT) \cite{VIT} S/16 is fine-tuned to encode the reflection maps into compact features capturing geometric structure and spatial relationships.
An overview of the geometry module is given in Appendix \labelcref{subsub: GeomModule}.

\subsection{Diffusion Transformer} \label{subsec: DiT}
Inspired by recent advances in image and audio generation \cite{peebles2023scalable, levy2023controllable, hung2024tangoflux, stableaudio, stableaudioopen}, we parameterize the velocity field $\widehat{\mathbf{v}}_t$ using a diffusion transformer (DiT) illustrated in \cref{fig: FS_DiT}.
It consists of a multi-layer Transformer architecture. A 1D convolution followed by a linear layer maps between the VAE latent space and the transformer embedding dimension. Each transformer block follows a fixed sequence: self-attention with Rotary Positional Embedding (RoPE) \cite{rope}, followed by cross-attention over conditioning tokens and a feedforward network (FNN), with residual connections applied inside each sub-layer. We compute $d$-dimensional Fourier features of the noise timestep $t$. 
The global conditioning, containing the target pose and $t$, is incorporated via Adaptive Layer Norm (AdaLN), where learned scale, shift and gating parameters modulate both self-attention and feedforward layers. Acoustic, spatial and geometric context are incorporated through cross-attention. 
Finally, the model consists of 12 transformer blocks with 8 heads and a hidden width of 256. We train it with a learning rate of $5 \times 10^{-5}$, AdamW optimizer \cite{AdamW} and a batch size of 64 on a single H100 GPU. We use an Exponential Moving Average (EMA) of the model weights during training and BF16 precision. \methodname~number of parameters and inference time are reported in the Appendix \labelcref{sec: SM_param}.

\begin{figure}[t]
    \centering
    \includegraphics[width=\linewidth]{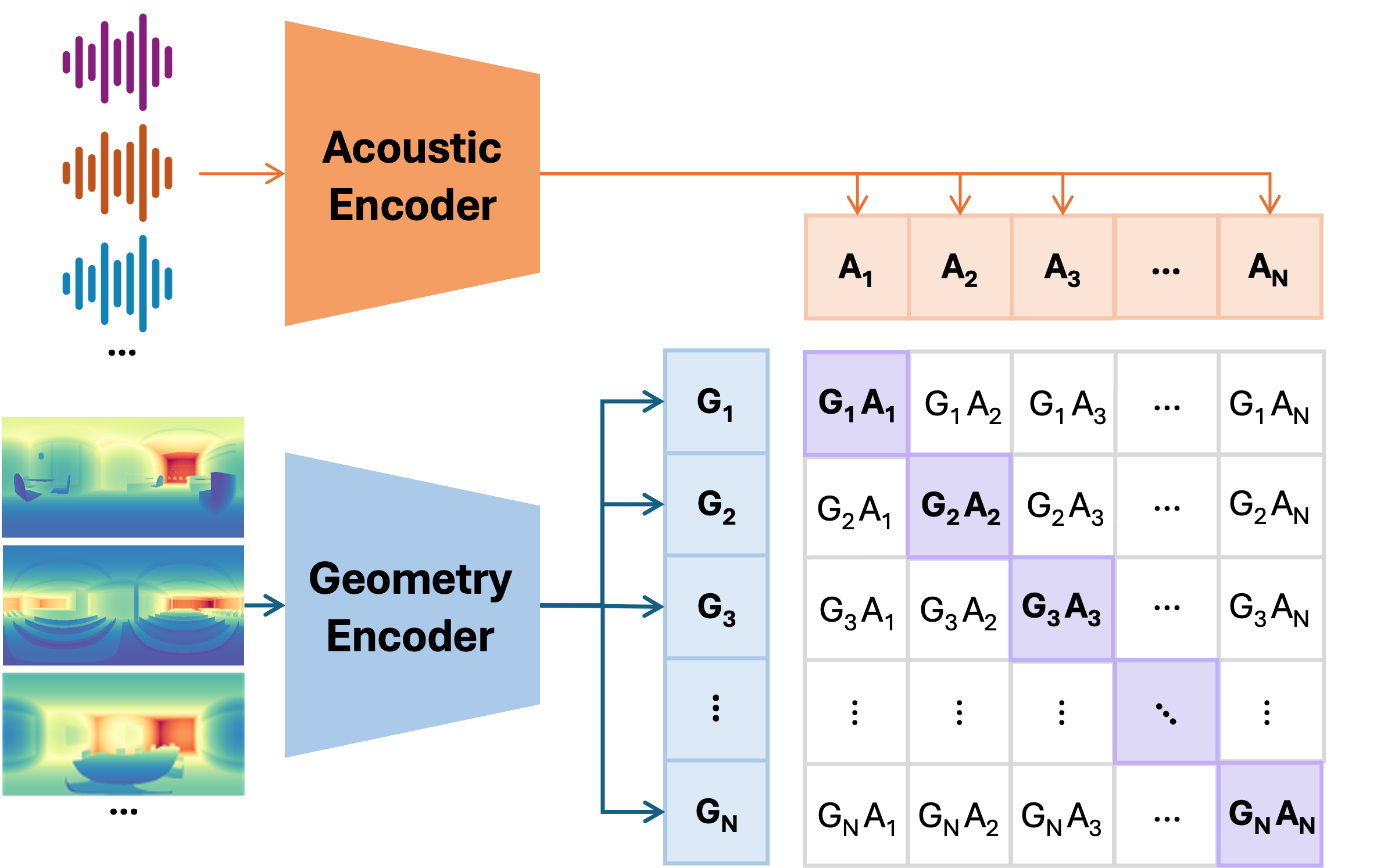}
    \caption{\textbf{AGREE contrastive framework:} Audio and geometry inputs are encoded into a shared latent space, where a contrastive objective maximizes similarity for matching pairs (diagonal entries) and minimizes it for mismatched ones.}
    \label{fig:acoustic-geometry}
\end{figure}

\section{AGREE: Acoustic–Geometry Embedding}
We introduce \AGname~(\underline{A}coustic-\underline{G}eomet\underline{R}y \underline{E}mb\underline{E}ding), a CLIP-style~\cite{CLIP} multimodal embedding that aligns room acoustics and geometry (see \cref{fig:acoustic-geometry}).
The audio encoder is fine-tuned from the pre-trained VAE encoder used in \methodname~(see \cref{subsec: VAE}). 
The geometry encoder is DINOv3 ViT-S/16 fine-tuned on panoramic depth maps captured at receiver positions. Following \methodname's geometry pipeline, depth values are projected into 3D coordinates and source positions, expressed in the receiver frame, are subtracted. 
Each encoder is followed by a linear projection, and both are trained jointly with a contrastive objective to align acoustic and geometric representations.
The resulting embedding space captures spatial-acoustic consistency, enabling zero-shot cross-modal retrieval and geometry-aware evaluation. 
AGREE details are provided in Appendix \labelcref{sec: SM_AGREE}.
\section{Experiments}
\begin{table*}[t]
\centering

\caption{\textbf{Performance on unseen AcousticRooms scenes:} Results are shown for $K\in\{8,1,$\,\xmark$\}$ reference RIRs. For \methodname, we report mean and standard deviation over 5 generations. 
\methodname~outperforms all baselines even in the one-shot setting. \abl~denotes ablations with either geometry (G) or audio conditioning removed.}

\newcolumntype{P}[1]{>{\hspace{0pt}}c<{\hspace{#1}}}
\resizebox{\linewidth}{!}{
    \begin{tabular}{l|c|c|P{2pt}P{5pt}P{2pt}cccc} %c}
    \toprule
    
     \textbf{Method} & \textbf{K} & \textbf{G} & \textbf{T60 (\%) $\downarrow$} & \textbf{C50 (dB) $\downarrow$} & \textbf{EDT (ms) $\downarrow$} & \textbf{R@1 (\%) $\uparrow$} & \textbf{R@5 (\%) $\uparrow$} & \textbf{R@10 (\%) $\uparrow$} & \textbf{$\text{FD}_G$ $\downarrow$} \\
      
    \midrule
    Random Across Rooms & \xmark & \xmark &
    44.73 & 7.676 & 306.29 & 0.02 & 0.06 & 0.32 & 0.111 \\ 
    Random Same Room  & \xmark & \xmark & 
    \textbf{17.36} & 5.490 & 168.17 & 0.25 & 1.09 & 2.16 & \textbf{0.001} \\ 
    \methodname\abl & \xmark & \checkmark &
    23.41\unc{0.02} & \textbf{2.554}\unc{0.002} & \textbf{109.75}\unc{0.09} & \textbf{5.12}\unc{0.12} & \textbf{16.47}\unc{0.14} & \textbf{23.12}\unc{0.14} & 0.337 \\ 

    \midrule 
    Nearest Neighbor & 1 & \xmark &
    15.22 & 5.212 & 157.94 & 0.00 & 2.26 & 4.56 & \textbf{0.001} \\ 
    Fast-RIR & 1 & \checkmark & 18.97 & 3.257 & 121.21 & 0.17 & 0.66 & 1.64 & 0.456 \\
    xRIR & 1 & \checkmark &
    14.47 & 1.961 & 74.45 & 0.28 & 1.36 & 2.59 & 0.263 \\ 
     \rowcolor{babyblue} \textbf{\methodname} & 1 & \checkmark & 
     \textbf{9.95}\unc{0.05} & \textbf{1.046}\unc{0.002} & \textbf{40.04}\unc{0.22} & \textbf{6.80}\unc{0.11} & \textbf{18.92}\unc{0.10} & \textbf{26.87}\unc{0.19} & 0.303 \\ 

    \midrule
    Linear Interpolation & 8 & \xmark &
    14.45 & 3.503 & 114.27 & 0.41 & 2.30 & 4.02 & 0.401 \\ 
    Nearest Neighbor & 8 & \xmark &
    10.91 & 2.792 & 90.08 & 0.00 & 10.26 & 17.28 & \textbf{0.003} \\ 
    \methodname\abl & 8 & \xmark & 
    12.07\unc{0.01} & 4.296\unc{0.001} & 140.04\unc{0.04} & 0.09\unc{0.01} & 0.58\unc{0.06} & 1.06\unc{0.04} & 0.663 \\
    Fast-RIR & 8 & \checkmark & 17.71 & 3.253 & 121.21 & 0.24 & 0.99 & 1.88 & 0.465 \\
    xRIR & 8 & \checkmark &
    9.98 & 1.354 & 49.40 & 0.54 & 2.00 & 3.38 & 0.307 \\ % epoch 36
    \rowcolor{babyblue} \textbf{\methodname} & 8 & \checkmark &
    \textbf{8.60}\unc{0.01} & \textbf{0.970}\unc{0.002}  & \textbf{37.13}\unc{0.02} & \textbf{6.99}\unc{0.13} & \textbf{19.38}\unc{0.15} & \textbf{27.21}\unc{0.17} & 0.305 \\ 
   
    \bottomrule
    \end{tabular}
}
\vspace{-12pt}

\label{tab: FS_AR_unseen}
\end{table*}

\subsection{Datasets} \label{subsub: Datasets}
\paragraph{AcousticRooms.}  
We use the AcousticRooms (AR) dataset \cite{liu2025haae}, a large-scale simulated dataset of monaural RIRs paired with equirectangular panoramic depth maps. It spans 260 rooms across 10 categories with diverse geometries, sizes, and materials, totaling over 300k simulated RIRs at 22,050\,Hz. Generated with Treble Technology’s wave-based simulation, it provides high simulation accuracy beyond geometric or ray-tracing methods used in \cite{chen20soundspaces, chen22soundspaces2, tang2022gwa}.
Following \cite{liu2025haae}, we split the dataset into 243 seen and 17 unseen rooms to evaluate both in-room prediction and generalization to new scenes. The unseen test set contains 5,244 instances. A subset of the seen-room instances is used for evaluation, it contains 6,217 instances across 131 rooms. In all our experiments, our VAE is pretrained on this dataset.

\vspace{-8pt}
\paragraph{Hearing-Anything-Anywhere.}
To evaluate generalization to real-world environments, we use the Hearing-Anything-Anywhere (HAA) dataset \cite{HAA}. It provides monaural RIRs recorded in four rooms, each with a fixed source and multiple receiver positions. This setup is the inverse of AcousticRooms, where the receiver is fixed and the source varies. However, for single-channel RIRs, interchanging source and receiver is equivalent due to the symmetry of the wave equation \cite{liu2025haae}. All RIRs are sampled to 22,050\,Hz. Panoramic depth maps at each source pose are derived from room meshes reconstructed using wall and surface annotations. Appendix \labelcref{subsub: DatasetDetails} gives datasets details. 

\subsection{Metrics}
\paragraph{Perceptual metrics.}
We assess the perceptual quality of generated RIRs using standard acoustic metrics \cite{liu2025haae, fewshot, INRAS, NeRAF, AVNeRF} that correlate with human auditory perception. We report the relative T60 error, normalized by the ground-truth. T60 measures the reverberation time \ie, the duration for sound energy to decay by 60\,dB. We also compute the clarity error based on C50, the ratio of early-to-late energy, indicative of speech intelligibility and acoustic clarity. Finally, we evaluate the Early Decay Time (EDT) error, which capture early reflection characteristics by measuring the time for an initial 5\,dB energy decay.

\vspace{-8pt}
\paragraph{Scene-consistency metrics.}
We introduce metrics based on the \AGname~embedding space to evaluate how well generated RIRs reflects the spatial characteristics of the environment.
We compute audio-to-audio recall (R@1/5/10), quantifying how closely generated and ground-truth RIRs align in this geometry-aware space. To capture overall realism, we compute the Fréchet distance $\text{FD}_G$, between the distribution of generated and real audio embeddings in AGREE space, analogous to the FID \cite{heusel2017gans} used in image generation.
For reference, AGREE zero-shot retrieval results on the unseen AcousticRooms set are summarized in \cref{tab: AGREE_summary_unseen}. 
To maximize retrieval performance when evaluating RIR synthesis methods, we also train AGREE on the entire AR dataset. Further details on the scene-consistency metrics can be found in Appendix \labelcref{sub: AGREEMetrics}.

\begin{table}[t]

\centering
\caption{\textbf{Zero-shot cross-modal retrieval on the unseen AcousticRooms set:} We report acoustic-to-geometry (A2G) and geometry-to-acoustic (G2A) recall at 1, 5 and 10. \dag~indicates training on the full dataset for benchmarking few-shot methods.
}
 
\resizebox{\linewidth}{!}{
    \begin{tabular}{l|ccc|ccc}
    \toprule

    \multirow{2}{*}{\textbf{Method}} & \multicolumn{3}{c|}{\textbf{A2G}} & \multicolumn{3}{c}{\textbf{G2A}} \\
     
     & \textbf{R@1$\uparrow$} & \textbf{R@5$\uparrow$} &\textbf{R@10$\uparrow$} &  \textbf{R@1$\uparrow$} & \textbf{R@5$\uparrow$} &\textbf{R@10$\uparrow$} \\
      
    \midrule

    AGREE & 59.78 & 83.53 & 89.35 & 59.10 & 85.56 & 91.04  \\
    AGREE$^{\dag }$ & 85.37 & 99.70 & 99.98 & 84.38 & 99.53 & 99.97 \\
  
    \bottomrule
    \end{tabular}
}
\vspace{-7.5pt}
\label{tab: AGREE_summary_unseen}
\end{table}

\subsection{Baselines}
We compare \methodname~against several baselines:
\begin{itemize}
    \item Random Across Rooms: randomly samples a RIR from the entire dataset.
    \item Random Same Room: randomly selects a RIR from the same room.
    \item Linear Interpolation: linearly interpolates $K$ reference RIRs based on their distances to the target source.
    \item Nearest Neighbor (KNN): chooses the RIR closest in distance to the target source among the $K$ references.
    \item Fast-RIR \cite{FastRIR}: generates RIRs with a GAN conditioned on T60 and scene size estimated from $K$ RIRs and depth.
    \item xRIR \cite{liu2025haae}: combines acoustic and geometric features to weight $K$ reference RIRs.
\end{itemize}

\subsection{Inference parameters}
\begin{figure}[t]
    \centering
    \includegraphics[width=\linewidth]{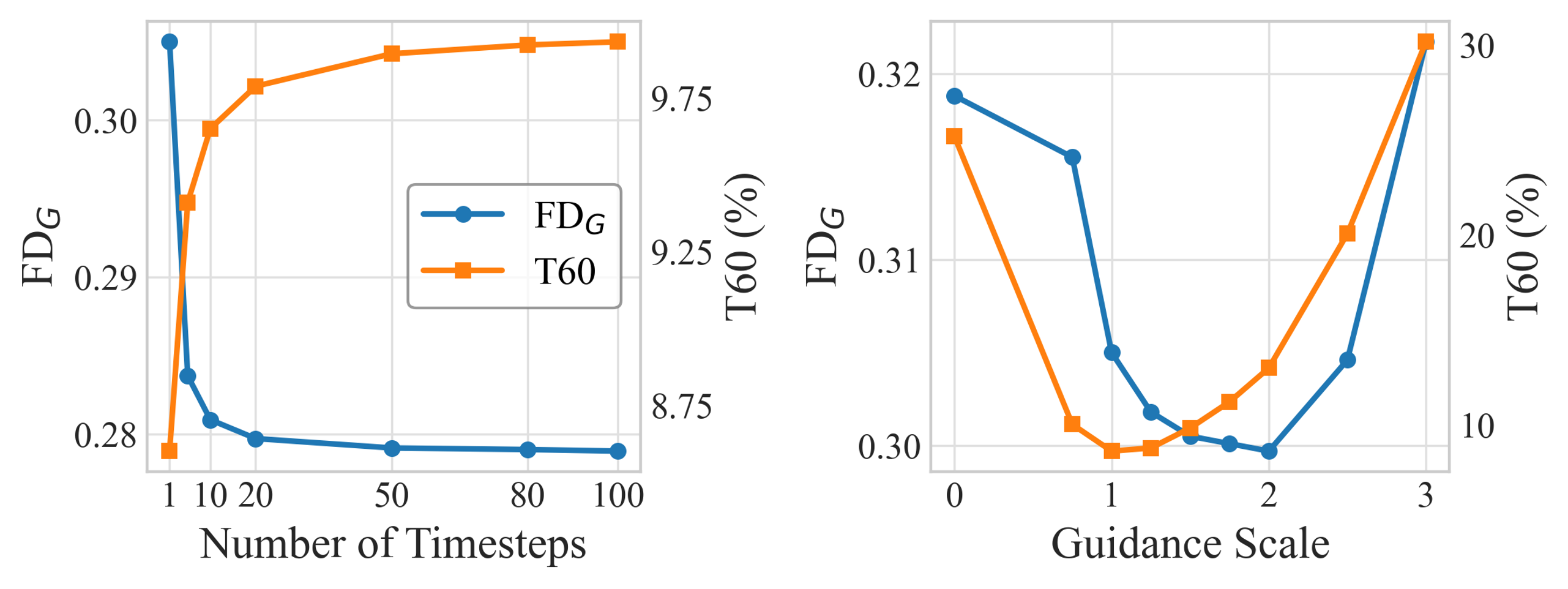}
    \caption{\textbf{Impact of classifier-free guidance and inference steps:} Evolution of T60 and FD$_G$ as a function of the guidance scale $\omega$ and the number of timesteps.}

    \label{fig:cfg-steps}
    \vspace{-15pt}
\end{figure}

In all experiments, we use a guidance scale of 1 and perform generation in a single inference step as it achieves the best results on the perceptual metrics (T60, C50, EDT). These metrics mainly capture global acoustic properties, such as energy decay and clarity, but are insensitive to fine-grained details or sample diversity. Thus, additional steps offer no benefit. However, as shown in \cref{fig:cfg-steps}, increasing the guidance weight or the number of steps improves FD$_G$.

\subsection{Results}
\paragraph{8-shot generation in novel environments.}
Quantitative results on unseen scenes with $K{=}8$ are reported in \cref{tab: FS_AR_unseen}. 
\methodname~consistently outperforms xRIR across all metrics, reducing errors by $13.8\%$ T60, $28.3\%$ C50, and $24.9\%$ EDT, and achieving higher audio-to-audio recall, indicating more geometry-consistent acoustic synthesis. For FD$_G$, \methodname~slightly surpasses xRIR, reflecting improved distributional realism. Increasing the number of inference steps or the classifier-free guidance weight further improves \methodname~FD$_G$; for example, 20 steps reduce it to 0.280 (see \cref{fig:cfg-steps}). KNN always achieves lower FD$_G$ as it simply returns a reference RIR, which is already drawn from the true distribution. 
For seen rooms, detailed results are provided in Appendix \labelcref{sub: seenset}: \methodname~reduces errors by $23.9\%$, $29.8\%$, and $24.8\%$ for T60, C50, and EDT, respectively.
These results demonstrate that \methodname~not only improves RIR estimation at new positions within seen spaces but also generalizes more effectively to new environments.

\vspace{-8pt}

\begin{figure}
    \centering
    \includegraphics[width=\linewidth]{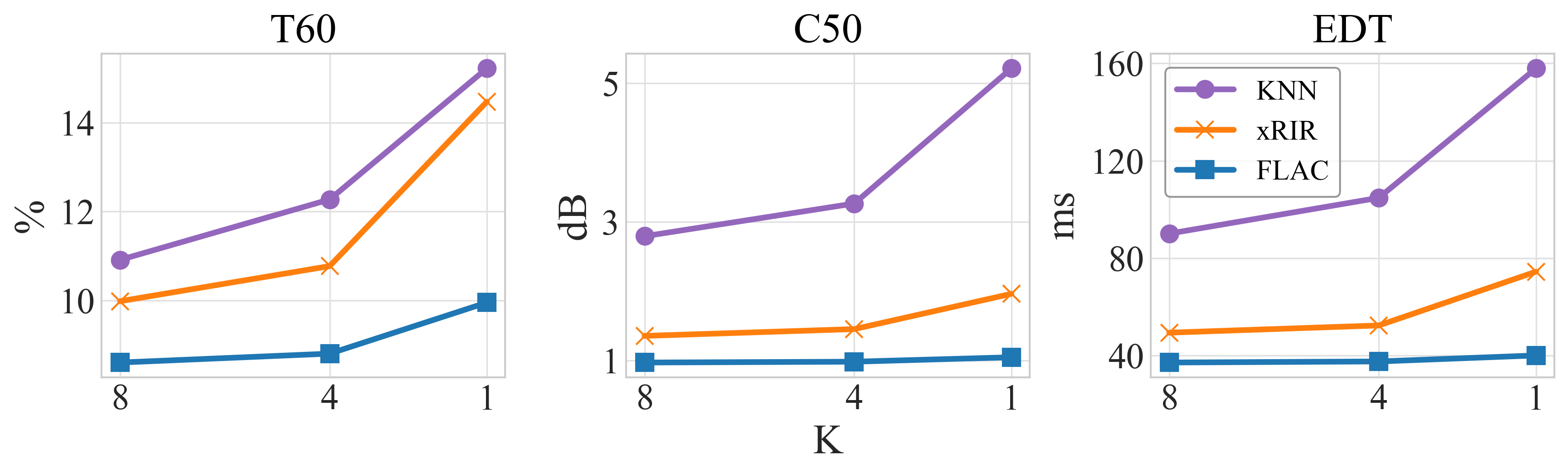}
    \caption{\textbf{Robustness to limited context RIRs in novel scenes:} Performance as the number of reference RIRs ($K$) decreases for KNN, xRIR, and \methodname. \methodname~remains the most stable and outperforms state-of-the-art methods with $K{=}8$ even in one-shot.}
    
    \label{fig: FS_RobustnessK}
    \vspace{-5pt}
\end{figure}

\paragraph{Robustness under limited observations.}  
We evaluate methods robustness with fewer context RIRs, simulating scenarios with limited recordings. For \methodname~and xRIR, models trained with $K{=}8$ and tested with fewer references.
As shown in \cref{tab: FS_AR_unseen}, \methodname~maintains state-of-the-art performance in the one-shot setting, surpassing prior methods using eight recordings. \cref{fig: FS_RobustnessK} shows that \methodname~remains significantly more stable than KNN and xRIR as $K$ decreases. 
Recall metrics are little affected by reduced acoustic observations, indicating that geometry provides the dominant cue for geometry-consistent RIR synthesis. 

\begin{figure}
    \centering
    \includegraphics[width=\linewidth]{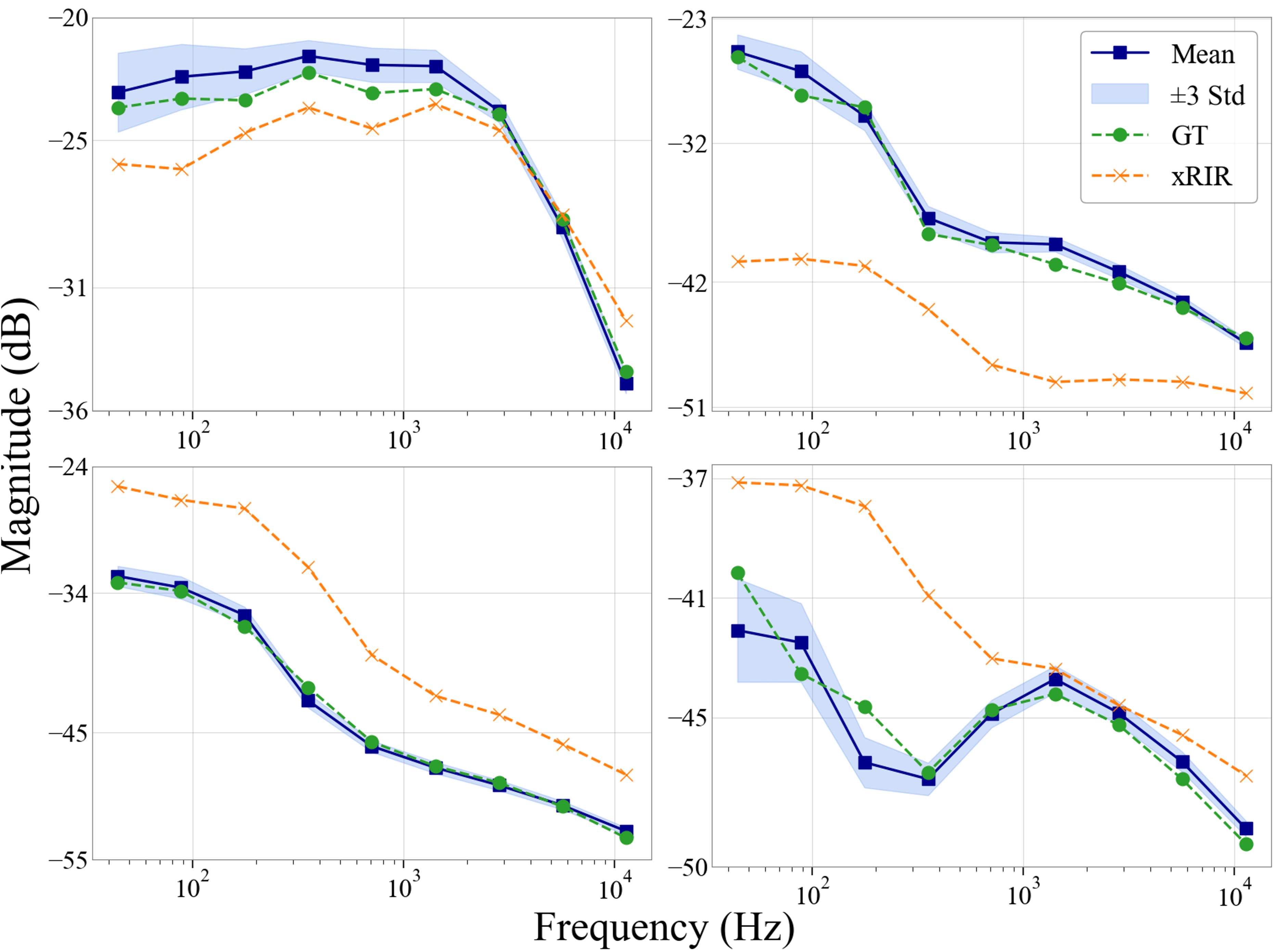}
    \caption{\textbf{Octave-band analysis of 4 RIRs in unseen rooms:}
    100 samples per instance are generated with \methodname. The mean and $\pm3$ standard deviation (covering 99.7\% of the distribution) are shown. Std increases at low frequencies.}
    
    \label{fig: FS_RIRDistribution}
    \vspace{-12pt}
\end{figure}

\begin{figure}
    \centering
    \includegraphics[width=0.7\linewidth]{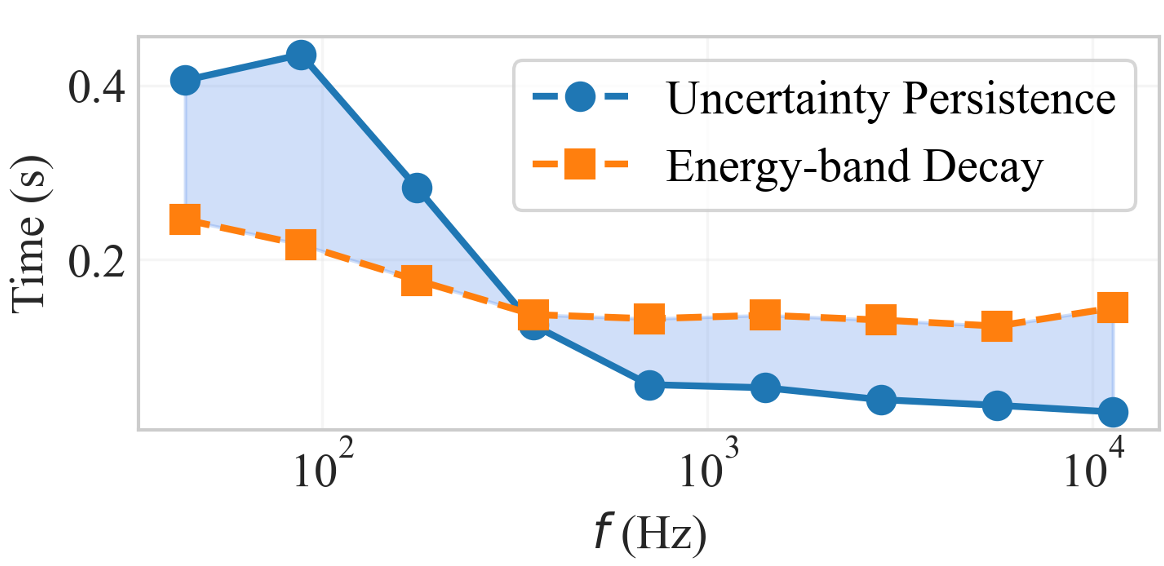}
    \vspace{-5pt}
    \caption{\textbf{Uncertainty persistence time} and band-wise energy decay, averaged over 100 unseen samples. Uncertainty lasts longer at low frequencies and decays faster at high frequencies.
    }
    \label{fig:uncertaintypersistence}
    
    \vspace{-12pt}
\end{figure}

\vspace{-8pt}
\paragraph{Capturing uncertainty of few-shot RIR synthesis.}
We study variability by generating 100 RIRs per conditioning, each produced with a different noise input. As shown by the octave-band analysis in \cref{fig: FS_RIRDistribution}, samples standard deviation increases at low frequencies. These bands also exhibit longer uncertainty persistence time, defined as the time until band-wise sample variance drops below the 75th percentile (see \cref{fig:uncertaintypersistence}). This matches room acoustics theory: low-frequency responses are governed by sparse, boundary-dependent modes that are weakly constrained by limited context, whereas above the Schroeder frequency dense mode yields stable responses constrained by local geometry. This indicates that \methodname~captures the inherent uncertainty of underconstrained few-shot settings. A deterministic variant (fixed noise) degrades performance (+6\% T60, +10\% C50, -40\% R@5), confirming that stochasticity is essential.
Quantitatively, \methodname's intra-conditioning diversity is $1.03_{\pm0.20}$ vs. $22.96$ between conditionings (a 4.5\% ratio), showing that \methodname~introduces meaningful stochasticity while remaining consistent with the context. See  Appendix \labelcref{sec: SM_quali} for a t-SNE visualization.

\begin{table}[t]
\centering
\caption{\textbf{Sim-to-real transfer to the Hearing-Anything-Anywhere dataset: } Few-shot methods are compared against Diff-RIR \amandine{and INRAS}, which require per-scene training ($^\dag$). For \methodname, we report mean and standard deviation over 5 generations. With $K{=}8$, \methodname~matches or exceeds xRIR and Diff-RIR on perceptual metrics, and with one-shot, it outperforms KNN and xRIR.}

\resizebox{\linewidth}{!}{

    \newcolumntype{P}[1]{>{\hspace{0pt}}c<{\hspace{#1}}}
    \begin{tabular}{l|c|P{2pt}P{5pt}P{2pt}cc}
    \toprule
    \textbf{Method} & \textbf{K} & \textbf{T60 (\%)} $\downarrow$ & \textbf{C50 (dB)} $\downarrow$ & \textbf{EDT (ms)} $\downarrow$ & \textbf{R@5 (\%)} $\uparrow$ & \textbf{FD}$_G$ $\downarrow$ \\
    
    \midrule
    Random Across Rooms & \xmark & 17.40 & 10.283 & 533.99 & 1.49 & 0.460 \\
    Random Same Room & \xmark & 8.00 & 4.805 & 180.15 & 1.86 & 0.169 \\

    \midrule
    Nearest Neighbor & 1 & 8.19 & 5.000 & 187.55 & 1.20 & \textbf{0.177} \\
    xRIR & 1 & 8.63 & 4.862 & 183.27 & 14.85 & 0.363 \\
    
    \rowcolor{babyblue} \textbf{\methodname} & 1 & \textbf{3.45}\unc{0.02} & \textbf{2.170}\unc{0.014} & \textbf{90.02}\unc{0.24} & \textbf{17.94}\unc{0.62} & 0.564 \\ 

    \midrule
    Linear Interpolation & 8 & 4.12 & 2.695 & 88.19 & 3.62 & 0.904 \\
    Nearest Neighbor & 8 & \textbf{2.89} & \textbf{1.923} & \textbf{77.24} & 9.61 & \textbf{0.169} \\
    xRIR & 8 & 6.53 & 3.492 & 149.69 & \textbf{20.65} & 0.318  \\ 
    \rowcolor{babyblue} \textbf{\methodname}  & 8 & 3.10\unc{0.01} & 2.167\unc{0.004} & 84.52\unc{0.24} & 17.41\unc{0.59} & 0.585 \\

    \midrule
    INRAS$^{\dag}$ & 12 & 6.61 & 3.966 & 158.07 & 2.27 & 0.797 \\
    Diff-RIR$^{\dag}$ & 12 & 3.74 & 2.067 & 88.09 & 26.97 & 0.263 \\ % max len 44100

    \bottomrule
    \end{tabular}
}

\label{tab: FS_HAA}
\vspace{-5pt}
\end{table}

\vspace{-8pt}
\paragraph{Sim-to-real transfer.}
We evaluate real-world generalization on the HAA dataset \cite{HAA}. Baselines include Diff-RIR \cite{HAA}, a physics-based differentiable renderer, and INRAS \cite{INRAS}, a novel-view acoustic synthesis method, both trained with 12 references per room to predict RIRs at new locations. Unlike few-shot models, they must be retrained separately for each room, requiring hours of training. Following \cite{liu2025haae}, we fine-tune xRIR and \methodname. Note that we do not fine-tune \methodname's VAE. Few-shot models adapt to all four rooms within minutes. For evaluation, AGREE is also fine-tuned on HAA.
As shown in \cref{tab: FS_HAA}, with eight shots, \methodname~outperforms xRIR and surpasses Diff-RIR on most perceptual metrics, despite using fewer references and no room-specific training. 
While 8-NN performs strongly, it copies existing RIRs, lacking spatial continuity (audible "jumps").
Remaining discrepancies likely stem from: (i) HAA’s simplified geometry annotations (\eg, tables as single planes); and (ii) the VAE not being fine-tuned on real recordings, which may cause its latent representation to miss certain acoustic phenomena. The small size of HAA proved insufficient for stable adaptation of the VAE. Yet, \methodname~one-shot outperforms both KNN and eight-shot xRIR, highlighting its advantage in data-scarce conditions.
%Note that inverting the receiver projection, as HAA setup is reversed, improves T60/C50/EDT but reduces geometry consistency (see Appendix).}

\vspace{-8pt}
\paragraph{Perceptual Evaluation.}
We conducted a listening study with 46 participants on 14 unseen AR scenes. Participants were presented with the ground-truth, audio generated by FLAC (1-shot) and xRIR (8-shot), and were asked to select which audio sounded closer to the GT. FLAC was preferred in 93.01\% of cases. Details are given in Appendix \labelcref{sec: SM_perceptual_eval}.

\subsection{Ablation Study} \label{ablation}
\begin{table}[t]
\centering
\caption{\textbf{Impact of geometry and acoustic encoders:} Performance on unseen AcousticRooms scenes using different configurations of the geometry $\phi_G$ and acoustic $\phi_A$ encoders. We compare xRIR’s ViT and DINOv3 ViT-S/16 with three initialization strategies: trained from scratch, frozen, or fine-tuned ($\mathcal{W}_{\text{DINO}}$). For $\phi_A$, we evaluate the ResNet-18 and our frozen VAE encoder.}

\resizebox{\linewidth}{!}{
    \begin{tabular}{cc|c|c|ccccc}
    \toprule
    \multicolumn{2}{c|}{\textbf{$\mathbf{\phi}_{\text{G}}$}} & \multirow{2}{*}{\textbf{$\mathbf{\phi}_{\text{A}}$}} & \multirow{2}{*}{\textbf{K}} & \multirow{2}{*}{\textbf{T60 (\%) $\downarrow$}} & \multirow{2}{*}{\textbf{C50 (dB) $\downarrow$}} & \multirow{2}{*}{\textbf{EDT (ms) $\downarrow$}} & \multirow{2}{*}{\textbf{R@5 (\%) $\uparrow$}} & \multirow{2}{*}{\textbf{FD$_G$ $\downarrow$}} \\

     \textbf{ViT} & \textbf{$\mathcal{W}_{\text{DINO}}$}  & & & & &  & & \\
    \midrule

    \cite{liu2025haae} & \xmark & ResNet & 1 & 10.91 & 1.166 & 42.41 & 11.17 & 0.328 \\ 
    
    S/16 \cite{dinov3} & \xmark & ResNet & 1 & 10.81 & 1.090 & 42.11 & 15.40 & 0.318 \\ % 52500
    S/16 \cite{dinov3} & \SnowflakeChevron & ResNet & 1 & 10.42 & 1.427 & 51.79 & 5.29 & 0.373 \\ 
    \rowcolor{babyblue} S/16 \cite{dinov3} & \checkmark & ResNet & 1 & 9.95 & \textbf{1.046} & 40.04 & \textbf{18.9} & \textbf{0.303}  \\ 
    
    S/16 \cite{dinov3} & \checkmark & VAE & 1 & \textbf{9.40} & 1.057 & \textbf{39.31} & 17.11 & 0.310 \\ 

    \midrule
    \cite{liu2025haae} & \xmark  & ResNet & 8 & 9.46 & 1.063 & 39.57 & 11.80 & 0.333 \\
    
    S/16 \cite{dinov3} & \xmark & ResNet & 8 & 9.29 & 0.994 & 38.61 & 16.24 & 0.320  \\ % 52500
    S/16 \cite{dinov3} & \SnowflakeChevron & ResNet & 8 & 8.87 & 1.298 & 46.41 & 5.92 & 0.378 \\ 
    \rowcolor{babyblue} S/16 \cite{dinov3} & \checkmark & ResNet & 8 & 8.60 & 0.970 & 37.13 & \textbf{19.38} & \textbf{0.305} \\ 
    
    S/16 \cite{dinov3}  & \checkmark & VAE & 8 & \textbf{8.51} & \textbf{0.945} & \textbf{34.70} & 17.56 & 0.310 \\
    
    \bottomrule
    \end{tabular}
}

    \vspace{-5pt}
\label{tab: EncoderAblation_unseen}
\end{table}
\begin{table}[t]
\centering
\caption{\textbf{Impact of DiT variants:} Performance on unseen AcousticRooms scenes with In-Context, Cross-Attention (CA), and hybrid AdaLN+CA conditioning.}

\resizebox{\linewidth}{!}{
    \begin{tabular}{l|c|ccccc}
    \toprule

      \textbf{Method} & \textbf{K} & \textbf{T60 (\%) $\downarrow$} & \textbf{C50 (dB) $\downarrow$} & \textbf{EDT (ms) $\downarrow$} & \textbf{R@5 (\%) $\uparrow$} & \textbf{FD$_G$ $\downarrow$}   \\

    \midrule
    In-Context & 1 & 69.68 & 11.199 & 1236.98 & 0.06 & 1.270 \\
    CA & 1 & 15.68 & 1.750 & 85.98 & 6.10 & 0.424 \\
    \rowcolor{babyblue} AdaLN+CA & 1 & \textbf{9.95} & \textbf{1.046} & \textbf{40.04} & \textbf{18.92} & \textbf{0.303} \\

    \midrule
    In-Context & 8 & \textbf{8.12} & 1.081 & 41.97 & 0.194 & 0.316 \\
    CA & 8 & 9.31 & 1.234 & 45.81 & 11.93 & 0.342 \\
    \rowcolor{babyblue} AdaLN+CA & 8 & \underline{8.60} & \textbf{0.970} & \textbf{37.13} & \textbf{19.38} & \textbf{0.305} \\

    \bottomrule
    \end{tabular}
}
			
    \vspace{-12pt}
\label{tab: Architecture_unseen}
\end{table}

\paragraph{Conditioning modalities.} \label{subsub: ablation_modalities}
We analyze the impact of each conditioning modality by removing either geometry or audio (see \cref{tab: FS_AR_unseen}).
When conditioned only on geometry, the model maintains strong audio-to-audio recall and outperforms random RIR prediction, confirming that geometric cues provide rich information for RIR synthesis. In contrast, using only audio leads to a drop in geometry-related metrics (recall and FD$_{G}$). %, with performance falling below KNN ($K=8$).
For perceptual metrics, geometry-only achieves higher C50 and EDT but lower T60 compared to the audio-only version. 
This aligns with their physical meaning: C50 and EDT are influenced by early reflections from nearby surfaces, while T60 captures global reverberation that is harder to infer from local geometry.  
Overall, combining both modalities through cross-attention yields the best results, demonstrating the complementary nature of geometric and acoustic conditioning.

\paragraph{Geometry conditioning encoder.}
In \cref{tab: EncoderAblation_unseen} we study how the choice of geometry conditioning encoder affects performance. We compare the ViT architecture from xRIR with DINOv3 ViT-S/16, which have similar parameter counts (19.8M vs.\ 21.7M). 
For DINOv3, we test three variants: (i) trained from scratch, (ii) frozen pretrained weights, and (iii) fine-tuned jointly with the model.
Even when trained from scratch, the ViT-S/16 outperforms xRIR’s ViT. As our input differs substantially from RGB images, freezing DINO weights degrades performance. Fine-tuning DINO yields the best overall results. 
We report a similar analysis for the AGREE geometric encoder in  Appendix \labelcref{subsub: AGREE_Encoder}, where fine-tuning DINOv3 ViT-S/16 consistently improves zero-shot retrieval.
Note that even with the same conditioning architecture as xRIR, one-shot \methodname~achieves comparable T60, FD$_G$ and higher C50, EDT and R@5 than eight-shot xRIR.

\vspace{-8pt}
\paragraph{Acoustic conditioning encoder.}
We evaluate replacing the jointly trained ResNet-18 with our frozen, pretrained VAE encoder (see \cref{tab: EncoderAblation_unseen}). The VAE improves cross-room generalization, though at higher computational cost. For efficiency, we use the ResNet-18 as the default encoder.% in our main experiments.

\vspace{-8pt}
\paragraph{DiT variants.} 
We study different DiT conditioning strategies (see \cref{tab: Architecture_unseen}).
\textit{In-Context} concatenates all conditioning information with the input before self-attention. \textit{Cross-Attention} applies conditioning solely via cross-attention layers. Our approach (see \cref{fig: FS_DiT}) injects target information through AdaLN, and contextual information via cross-attention. 
\textit{AdaLN+CA} outperforms alternative designs.
Illustrations of variants are provided in Appendix \labelcref{subsub: DiTVariants}.

\vspace{-3pt}

\section{Conclusion}
We introduced \methodname, a generative approach for few-shot acoustic synthesis based on flow matching. By conditioning generation on multimodal few-shot context, \methodname~can synthesize RIRs at arbitrary sensor positions in novel environments. Our method captures the inherent ambiguity of few-shot RIR synthesis, an aspect overlooked by existing deterministic methods. Experiments on two datasets demonstrated state-of-the-art performance in novel environments, even with a single reference RIR. We also introduced AGREE, a joint-embedding space between RIRs and geometry enabling both zero-shot cross-modal retrieval and geometry-consistency evaluation. \methodname~produces RIRs that are both perceptually accurate and consistent with the scene, an important aspect for immersive virtual experiences.
Future work may include supporting multiple sample rates in a single model, and collecting a larger, more  diverse real-world audio-visual dataset to improve sim-to-real transfer. The AGREE embedding could also benefit broader audio-visual learning tasks.

\section*{Acknowledgments}
This work was supported by the French Agence Nationale de la Recherche (ANR), under grant ANR22-CE94-0003 and was granted access to the HPC resources of IDRIS under the allocation 2024-AD011015475R1 made by GENCI. We would like to thank Simon de Moreau and the anonymous reviewers for their insightful comments and suggestions. %\newpage

{
    \small
    \bibliographystyle{ieeenat_fullname}
    \bibliography{main}
}

\clearpage
\setcounter{page}{1}
\setcounter{table}{0}
\setcounter{figure}{0}
\setcounter{section}{0}
\maketitlesupplementary

\appendix % For letters in section

% Redefine figure numbering
\renewcommand{\thefigure}{A.\arabic{figure}}
\renewcommand{\thetable}{A.\arabic{table}}

In the supplementary material, we first present details of our VAE (\cref{sec: SM_VAE}), FLAC (\cref{sec: SM_FLAC}), and AGREE (\cref{sec: SM_AGREE}). We then provide additional information on our evaluation setup, including datasets, metrics, and baselines (\cref{sec: SM_Eval}). Further results are reported in \cref{sec: SM_Results}, such as performances on the seen set of AcousticRooms, on HAA with a reversed setup, and complementary evaluation metrics. Additional qualitative examples, including a video, are provided in \cref{sec: SM_quali}. We give details about the perceptual evaluation setup in \cref{sec: SM_perceptual_eval}. Finally, we provide the number of parameters and inference speed of models in \cref{sec: SM_param}.

\section{VAE} \label{sec: SM_VAE}
\subsection{Training objective} \label{sub: VAE_detailed_training_obj}
We provide details on each loss used to train the VAE. In the following, let $\mathbf{x}$ and $\hat{\mathbf{x}}$ denote the ground truth and predicted waveforms and $\mathbf{X}$, $\widehat{\mathbf{X}}$ the magnitudes of their STFT representations.

We employ a multiresolution STFT loss $\mathcal{L}_{\text{MR}}$ inspired by \cite{steinmetz2020auraloss, yamamoto2020parallel}. This loss compares the spectrograms of the ground-truth and predicted waveforms at $m$ different resolutions using the spectral convergence $\mathcal{L}_{\text{SC}}$, log-magnitude $\mathcal{L}_{\text{SL}}$ and energy-decay losses $\mathcal{L}_{\text{ED}}$: 
\begin{multline}
    \mathcal{L}_{\text{MR}} (\mathbf{x}, \hat{\mathbf{x}}) = \sum_{i=1}^{m} \mathcal{L}_{\text{SC}} (\mathbf{X}_i, \widehat{\mathbf{X}}_i) \\ + \mathcal{L}_{\text{SL}} (\mathbf{X}_i, \widehat{\mathbf{X}}_i)  + \mathcal{L}_{\text{ED}} (\mathbf{X}_i, \widehat{\mathbf{X}}_i),
\end{multline}

\noindent with
\begin{gather}
    \mathcal{L}_{\text{SC}} (\mathbf{X}_i, \widehat{\mathbf{X}}_i) = \frac{||\mathbf{X}_i - \widehat{\mathbf{X}}_i||_F}{||\mathbf{X}_i||_F}, \\
    \mathcal{L}_{\text{SL}} (\mathbf{X}_i, \widehat{\mathbf{X}}_i) = ||\log(\mathbf{X}_i + \eta) - \log({\widehat{\mathbf{X}}_i} + \eta)||_1, \\
    \mathcal{L}_{\text{ED}} (\mathbf{X}_i, \widehat{\mathbf{X}}_i) = || 10\log_{10} \text{E}(\mathbf{X}_i) - 10\log_{10} \text{E}(\widehat{\mathbf{X}}_i) ||_1,
\end{gather}

\noindent where
\begin{equation}
    \text{E}(\mathbf{X}_i) = \sum_{k=d}^{T} \sum_{f=1}^{F} (\mathbf{X}_i(f,k))^2, \ 1 \leq d \leq T.
\end{equation}

To further improve the quality of the generated samples, we employ an adversarial hinge loss $\mathcal{L}_{\text{adv}}$ and feature matching loss $\mathcal{L}_{\text{feat}}$, based on the multi-scale STFT discriminator from Encodec \cite{Encodec}. Multi-scale discriminators are well suited for capturing structures in audio signals \cite{kumar2019melgan, hifiGAN, you2021gan}. The discriminator consists of multiple identically structured networks operating on multi-scaled complex-valued STFT, with the real and imaginary parts concatenated. Each sub-network is composed of a 2D convolutional layer, followed by 2D convolutions with increasing dilation rates of 1, 2 and 4 in the time dimension and a stride of 2 along the frequency axis. A final 2D convolution with kernel size 3 $\times$ 3 and stride $(1, 1)$ produces the output prediction. We use 5 scales with STFT window lengths $[2048, 1024, 512, 256, 128]$, hop lengths $[512, 256, 128, 64, 32]$ and FFT sizes $[2048, 1024, 512, 256, 128]$. 

The losses are expressed as:
\begin{gather}
    \mathcal{L}_{\text{adv}}(\mathbf{x}, \hat{\mathbf{x}}) = \sum_{n=1}^{N}\max[0, 1-D_n(\mathbf{x})] + \max[0, 1+D_n(\hat{\mathbf{x}})], \\
    \mathcal{L}_{\text{feat}}(\mathbf{x}, \hat{\mathbf{x}})= \frac{1}{NL} \sum_{n=1}^{N} \sum_{l=1}^{L} \frac{||D_n^l(\mathbf{x}) - D_n^l(\hat{\mathbf{x}})||_1}{\text{mean}(||D_n^l(\mathbf{x})||_1)},
\end{gather}
where $D_n^l$ is the l-th layer of the n-th discriminator $D_n$.

Finally, we use a KL divergence loss $\mathcal{L}_{\text{KL}}$ to regularize the latent distribution. The KL loss is given by:
\begin{align}
    \mathcal{L}_{\text{KL}}(\mathbf{x}) 
    &= D_{\text{KL}}\!\big(q_E(\mathbf{z}|\mathbf{x}) \,\|\, \mathcal{N}(0, \mathbf{I})\big),% \\
\end{align}
where $q_E(\mathbf{z}|\mathbf{x})$ denotes the encoder’s approximate posterior, $\mu_j $ and $\sigma_j$ are the mean and standard deviation of the $j$-th dimension of the latent variable $z$ and $d$ is the latent dimensionality.

\subsection{Implementation details} \label{sub: VAE_implementation_details}
We train the VAE using the AdamW optimizer \cite{AdamW} with a batch size of 64 on a single H100 GPU. The generator is optimized with a learning rate of $1.5 \times 10^{-5}$, the discriminator uses $3 \times 10^{-5}$. For the loss terms, all components of the multiresolution STFT loss are equally weighted, the KL loss is weighted by $1 \times 10^{-4}$, the adversarial loss by $0.1$, and the feature matching loss by $5.0$. The code is based on the stable audio tools library \footnote{\url{https://github.com/Stability-AI/stable-audio-tools}}.

\begin{figure*}
    \centering
    \includegraphics[width=\linewidth]{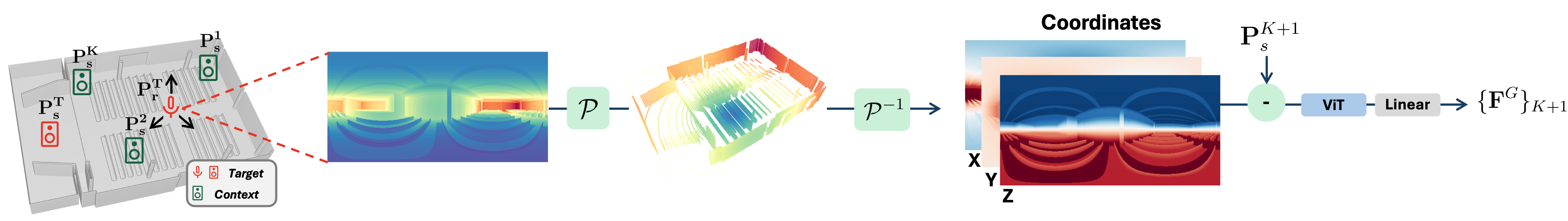}
    \caption{\textbf{Geometry module pipeline:} A panoramic depth map captured at the receiver position is unprojected into a 3D point cloud using the equirectangular projection $\mathcal{P}$, then reprojected so that each pixel encodes its corresponding 3D coordinates. We subtract the source pose associated with the RIR (previously projected in the receiver frame). The resulting representation is processed by a ViT followed by a linear layer to produce geometry-aware features. In HAA, the source and receiver are interchanged.}
    \label{fig:GeomModuleDino}
\end{figure*}

\section{FLAC} \label{sec: SM_FLAC}

\subsection{Implementation details}
When training on the synthetic dataset, we apply two forms of data augmentation to improve robustness: (i) a small random time shift of up to 10 samples in the time domain, and (ii) the addition of pink noise with a randomly chosen SNR between 40 and 60~dB.

Our DiT model consists of 12 transformer blocks with 8 heads and a hidden width of 256. We train it using a flow matching objective using a learning rate of $5 \times 10^{-5}$, AdamW optimizer \cite{AdamW} and a batch size of 64 on a single H100 GPU. We use an Exponential Moving Average (EMA) of the model weights during training and BF16 precision.

\subsection{Illustration of the geometry module} \label{subsub: GeomModule}
We illustrate the geometry module in \cref{fig:GeomModuleDino}.

\subsection{Illustration of the DiT variants} \label{subsub: DiTVariants}
\begin{figure*}
    \centering
    \includegraphics[width=\linewidth]{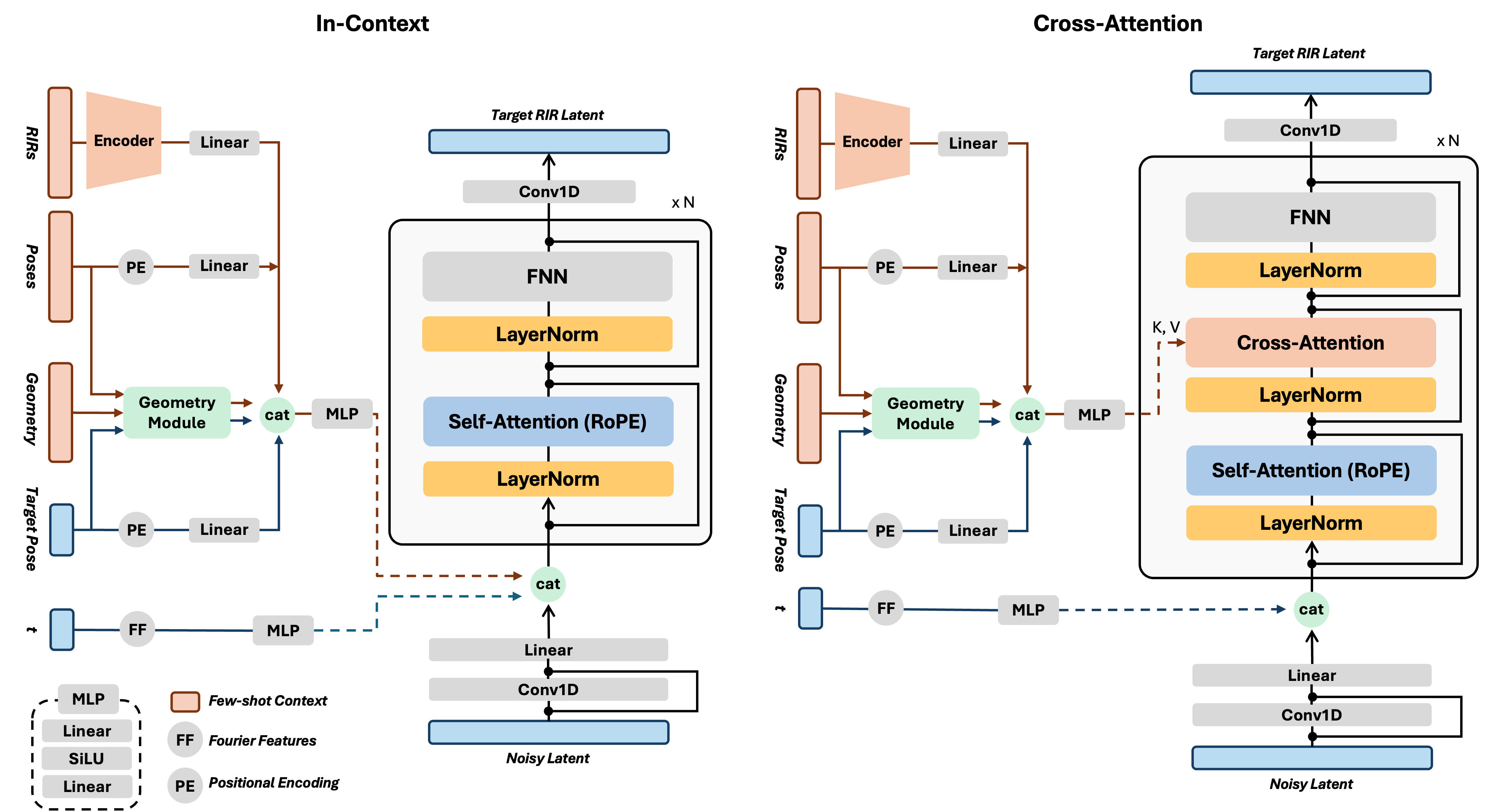}
    \caption{\textbf{DiT architecture variants:} \textit{In-Context} and \textit{Cross-Attention}–only architectures. The \textit{In-Context} variant concatenates all conditioning information with the input before self-attention. In the \textit{Cross-Attention} variant, conditioning is applied solely via cross-attention layers.}
    \label{fig: InCtxt_AllCA}
\end{figure*}

In \cref{ablation}, we compare our \textit{AdaLN+CA} DiT architecture (shown in \cref{fig: FS_DiT}) with two alternative designs: the \textit{In-Context} and the \textit{Cross-Attention} (CA-only) variants. Both alternative architectures are illustrated in \cref{fig: InCtxt_AllCA}.

\subsection{Conditioning: geometry and materials}
Panoramic depth maps do not recover occluded surfaces, creating ambiguity that directly motivates our stochastic formulation. While richer geometry (e.g., meshes) could reduce this, it introduces heavier assumptions. FLAC (1-shot) achieves strong performance, and depth can be estimated from RGB using foundation models, making an RGB + 1-RIR setup practical.

Similarly, while material-aware modeling have shown to improve prediction \cite{avrir, saad2025would}, explicit annotations are unavailable in real-world datasets like HAA. Instead, we rely on implicit cues in conditioning RIRs, shown to capture room-material properties \cite{liu2025haae}.

\section{AGREE} \label{sec: SM_AGREE}

\subsection{Training objective}
Given a batch $B$ of geometry embeddings $\mathbf{G} \in \mathbb{R}^{B \times d}$ and acoustic embeddings $\mathbf{A} \in \mathbb{R}^{B \times d}$, our model is trained using a symmetric contrastive objective analogous to CLIP \cite{CLIP}. We first compute pairwise similarity logits
\begin{equation}
    \mathbf{L}_{G} = \lambda \, \mathbf{G}\mathbf{A}^\top, \qquad 
    \mathbf{L}_{A} = \lambda \, \mathbf{A}\mathbf{G}^\top,
\end{equation}
where $\lambda$ is a learnable logit scaling parameter. 
Each row of $\mathbf{L}_{G}$ (resp.\ $\mathbf{L}_{A}$) contains the similarities between one geometry (resp.\ acoustic) embedding and all acoustic (resp.\ geometry) embeddings in the same batch. The ground-truth alignment corresponds to matching indices, $\mathbf{y} = (1,\dots,B)$, and the loss is defined as
\begin{equation}
    \mathcal{L}_{\text{contrast}} = \frac{1}{2}\Big(
    \text{CE}(\mathbf{L}_{G}, \mathbf{y}) + 
    \text{CE}(\mathbf{L}_{A}, \mathbf{y})
    \Big),
\end{equation}
where CE denotes cross-entropy loss.
It encourages aligned geometry-acoustic pairs to have high similarity while pushing apart mismatched pairs. 

\subsection{Implementation details}
AGREE operates on waveforms sampled at 22.05\,kHz of length 10,240, matching the training data from AcousticRooms \cite{liu2025haae} and HAA \cite{HAA}. 

For geometry, we use $256 \times 512$ panoramic depth maps captured at each receiver location in AcousticRooms (and at each source location in HAA). Following \methodname, depth maps are unprojected via equirectangular projection to obtain 3D points, and projected back in the image space so each pixel contains 3D coordinates. Then, we subtract the source pose expressed in the receiver frame (or the receiver pose for HAA). This provides the geometric context around the RIR’s recording configuration. \cref{fig:GeomModuleDino} illustrates this process.% in \methodname.

We jointly fine-tune DINOv3 ViT-S/16 \cite{dinov3} encoder and our frozen VAE audio encoder pretrained on AcousticRooms. A linear layer maps each encoder’s output into a shared 512-dimensional embedding space.

Training uses AdamW \cite{AdamW} with a learning rate of $1e^{-4}$, cosine decay, 10{,}000 warm-up steps, and a weight decay of 0.1. We employ a batch size of 128 and an embedding dimensionality of 512. The model is trained for 100 epochs. We base our implementations on OpenCLIP \cite{OpenCLIP} 
\footnote{\url{https://github.com/mlfoundations/open_clip}}.

\subsection{Impact of the geometry encoder} \label{subsub: AGREE_Encoder}
We compare several geometry encoder setup. Specifically, we compare four ViT variants: the encoder from \cite{liu2025haae}, the ViT-S/16 implementations from OpenCLIP \cite{OpenCLIP} and DINOv3, and the larger DINOv3 ViT-S+/16. For DINOv3, we assess the impact of using the pre-trained weights. Zero-shot retrieval results reported in \cref{tab: AGREE_full_unseen}.

With no specific weights initialization ($\mathcal{W}_{\text{DINO}} =$ \xmark), using the ViT-S/16 from DINOv3 achieves the best performance. Using frozen DINO weights is less effective, likely because our geometric representation differs significantly from the RGB images on which DINOv3 was pretrained. The strongest results are obtained by fine-tuning DINOv3 on our task, with ViT-S+/16 providing the best zero-shot retrieval on most metrics.

Since in this work AGREE is primarily used as a shared embedding space for evaluating the scene consistency of generated RIRs, we also train it on the full AcousticRooms dataset. In this setting, the performance gap between ViT-S and ViT-S+ becomes negligible, and we therefore adopt the smaller ViT-S/16 in the main experiments.

\subsection{AGREE vs. CRIP}
AGREE differs from CRIP \cite{avrir} in three aspects: (i) \textit{Local alignment}: CRIP uses RGB images unaligned with the RIR sensors, whereas AGREE uses pano depth captured at the receiver location. (ii) \textit{Early reflections}: AGREE encodes local surfaces that govern early reflections (while also capturing global structure), whereas CRIP primarily correlates with late reverberation; ablating comparable geometry in FLAC significantly degrades early-reflection metrics (C50, EDT, see \cref{tab: FS_AR_unseen}). (iii) \textit{Evaluation role}: CRIP is an auxiliary training signal, whereas AGREE serves as a geometry-aware evaluation framework.

\begin{table}[h]

\centering
\caption{\textbf{Zero-shot retrieval on the unseen split of the AcousticRooms dataset using different geometry encoders:} 
We report both acoustic-to-geometry (A2G) and geometry-to-acoustic (G2A) results for several ViT variants. 
$\mathcal{W}_{\text{DINO}}$ denotes DINOv3 pre-trained weights. 
Models marked with $^\dag$ are trained on the full AcousticRooms dataset.}

\resizebox{\linewidth}{!}{
    \begin{tabular}{cc|ccc|ccc}
    \toprule
      \multicolumn{2}{c|}{\textbf{$\mathbf{\phi}_{\text{G}}$}} & \multicolumn{3}{c|}{\textbf{A2G}} & \multicolumn{3}{c}{\textbf{G2A}} \\

      \textbf{ViT} & \textbf{$\mathcal{W}_{\text{DINO}}$} & \textbf{R@1$\uparrow$} & \textbf{R@5 $\uparrow$} & \textbf{R@10 $\uparrow$} & \textbf{R@1 $\uparrow$} & \textbf{R@5 $\uparrow$} & \textbf{R@10  $\uparrow$} \\
    \midrule
    \cite{liu2025haae} & \xmark & 6.26 & 17.82 & 25.69 & 6.63 & 19.20 & 27.30 \\
    S/16 \cite{OpenCLIP} & \xmark & 26.53 & 54.96 & 67.19 & 27.47 & 55.78 & 67.68  \\
    S/16 \cite{dinov3} & \xmark & 42.26 & 72.08 & 81.11 & 42.32 & 71.28 & 80.29 \\
    S/16 \cite{dinov3} & \SnowflakeChevron & 2.00 & 7.42 & 12.77 & 1.94 & 7.40 & 13.08 \\
    \rowcolor{babyblue} S/16 \cite{dinov3} & \checkmark & \textbf{59.78} & 83.53 & 89.35 & 59.10 & 85.56 & 91.04  \\
    S+/16 \cite{dinov3} & \checkmark & 58.17 & \textbf{85.06} & \textbf{90.72} & \textbf{60.01} & \textbf{85.85} & \textbf{92.00}    \\
    
    \midrule
    S/16 \cite{dinov3}$^\dag$ & \checkmark & 85.37 & 99.70 & 99.98 & 84.38 & 99.53 & 99.97  \\
    S+/16 \cite{dinov3}$^\dag$ & \checkmark & 85.26 & 99.68 & 99.97 & 84.28 & 99.78 & 99.98 \\
  
    \bottomrule
    \end{tabular}
}
			
\label{tab: AGREE_full_unseen}
\end{table}

\section{Evaluation details} \label{sec: SM_Eval}
\begin{table*}[t]
\centering

\caption{\textbf{Performance on seen AcousticRooms scenes:} Results are shown for $K\in\{8,1,$\,\xmark$\}$ reference RIRs. For \methodname, we report mean and standard deviation over 5 generations. 
\methodname~outperforms all baselines even in the one-shot setting. \abl~denotes ablations with either geometry (G) or audio conditioning removed.}

\resizebox{\linewidth}{!}{
    \begin{tabular}{l|c|c|P{2pt}P{5pt}P{2pt}cccc} 
    \toprule
     
     \textbf{Method} & \textbf{K} & \textbf{G} & \textbf{T60 (\%) $\downarrow$} & \textbf{C50 (dB) $\downarrow$} & \textbf{EDT (ms) $\downarrow$} & \textbf{R@1 (\%) $\uparrow$} & \textbf{R@5 (\%) $\uparrow$} & \textbf{R@10 (\%) $\uparrow$} & \textbf{$\text{FD}_G$ $\downarrow$}  \\
      
    \midrule
    Random Across Rooms & \xmark & \xmark &
    44.02 & 6.415 & 274.43 & 0.02 & 0.10 & 0.21 & 0.061 \\
    
    Random Same Room  & \xmark & \xmark & 
    \textbf{13.47} & 3.582 & 126.28 & 1.03 & 3.12 & 4.49 & \textbf{0.006} \\

    \methodname\abl & \xmark & \checkmark &
    24.56\unc{0.03} & \textbf{2.464}\unc{0.003} & \textbf{114.84}\unc{0.11} & \textbf{4.97}\unc{0.14} & \textbf{15.63}\unc{0.17} & \textbf{22.52}\unc{0.13} & 0.326 \\ 

    \midrule 
    Nearest Neighbor & 1 & \xmark &
    11.91 & 3.406 & 120.72 & 0.00 & 6.79 & 9.96 & \textbf{0.001} \\

    FastRIR & 1 & \checkmark & 15.88 & 2.152 & 92.44 & 0.23 & 0.89 & 1.64 & 0.383 \\

    xRIR & 1 & \checkmark &
    10.66 & 1.476 & 58.67 & 0.39 & 1.72 & 3.15 & 0.270 \\

     \rowcolor{babyblue} \textbf{\methodname}  & 1 & \checkmark & 
     \textbf{6.46}\unc{0.07} & \textbf{0.692}\unc{0.004} & \textbf{28.11}\unc{0.08} & \textbf{8.48}\unc{0.12} & \textbf{21.54}\unc{0.20} & \textbf{28.77}\unc{0.13} & 0.296 \\

    \midrule
    Linear Interpolation & 8 & \xmark &
    11.30 & 2.339 & 86.59 & 0.87 & 4.13 & 7.03 & 0.393 \\

    Nearest Neighbor & 8 & \xmark &
    8.31 & 1.802 & 64.77 & 0.08 & \textbf{23.08} & \textbf{30.71} & \textbf{0.002}\\

    \methodname\abl & 8 & \xmark & 
    10.47\unc{0.01} & 3.500\unc{0.001} & 125.99\unc{0.07} & 0.10\unc{0.01} & 0.41\unc{0.04} & 0.86\unc{0.04} & 0.644 \\

    FastRIR & 8 & \checkmark & 15.19 & 2.063 & 86.46 & 0.14 & 0.79 & 1.61 & 0.381 \\
    
    xRIR & 8 & \checkmark &
    6.99 & 0.916 & 34.15 & 0.48 & 2.38 & 4.02 & 0.328 \\  % epoch 36

    \rowcolor{babyblue} \textbf{\methodname} & 8 & \checkmark &
    \textbf{5.32}\unc{0.01} & \textbf{0.643}\unc{0.001} & \textbf{25.69}\unc{0.05} & \textbf{8.52}\unc{0.09} & 21.89\unc{0.20} & 29.28\unc{0.16} & 0.299 \\

    \bottomrule
    \end{tabular}
}

\label{tab: FS_AR_seen}
\end{table*}

\subsection{Datasets} \label{subsub: DatasetDetails}
\paragraph{AcousticRooms.}
Each RIR in AcousticRooms is sampled at 22.05\,kHz and truncated to 9,600 samples ($\approx$0.435\,s). We compute the magnitude spectrograms of the contextual RIRs before feeding them to the ResNet18 using an FFT size of 124, a window length of 62, and a hop size of 31. The panoramic depth maps are provided at a resolution of 256$\times$512 and are projected into 3D point clouds using equirectangular projection.

The dataset includes randomized material assignments drawn from 332 materials across 11 categories, ensuring strong diversity in acoustic behavior even among similar geometries. As the simulation meshes are untextured, no RGB data are available.

\paragraph{HAA.}
RIRs in the HAA dataset are originally sampled at 48\,kHz. We downsample them to 22.05\,kHz using Librosa's \texttt{resample} and truncate them to 9,600 samples to match our setup. Contextual RIRs are transformed using the same FFT pipeline as for AcousticRooms. 
While the dataset does not provide depth maps, simplified surface annotations are available\footnote{\url{https://github.com/maswang32/hearinganythinganywhere}}. We use these to reconstruct a mesh with Open3D \cite{open3D} and generate panoramic depth maps at each source position via Open3D raycasting. Note that, due to the simplified surface annotations, these depth maps differ substantially from those in AcousticRooms, widening the domain gap between the datasets. 
The test set comprises 1,282 instances (\textit{Base} rooms). To compute metrics, we first average results within each room and then average across the four rooms, preventing rooms with more data from dominating the results. Following \cite{liu2025haae}, we exclude the mean T60 for the dampened room, as all methods report unusually high values, likely due to its particular acoustic characteristics.

\subsection{Perceptual metrics}
Following \cite{liu2025haae}, we compute metrics on waveform of length 8,000 on the AcousticRooms dataset and 9,600 on the HAA dataset.
We use T60, C50 and EDT errors obtained as follows, where $\hat{\mathbf{x}}$ is the synthesized RIR waveform:
\begin{gather}
    \text{T60}(\hat{\mathbf{x}}, \mathbf{x}) = \frac{|\text{T60}(\hat{\mathbf{x}}) - \text{T60}(\mathbf{x})|}{\text{T60}(\mathbf{x})}, \\
    \text{C50}(\hat{\mathbf{x}}, \mathbf{x}) = |\text{C50}(\hat{\mathbf{x}}) - \text{C50}(\mathbf{x})|, \\
    \text{EDT}(\hat{\mathbf{x}}, \mathbf{x}) = |\text{EDT}(\hat{\mathbf{x}}) - \text{EDT}(\mathbf{x})|.
\end{gather}
For the AcousticRooms dataset, T60 is estimated based on T20, fitting the decay between -5\,dB and -25\,dB and linearly extrapolating to 60\,dB. For HAA, it is based on T30, following \cite{liu2025haae} implementation.

\subsection{Scene-consistency metrics} \label{sub: AGREEMetrics}
To evaluate how well generated RIRs preserve geometry-consistent acoustic behavior, we use AGREE, a CLIP-style audio-geometry joint embedding network. Let $\phi_A$ and $\phi_G$ denote its acoustic and geometry encoders, mapping audio $\mathbf{x}$ and geometry $g$ into a shared space. We propose metrics that both measure instance-level alignment (recall) and global distributional consistency (Fréchet Distance).

\paragraph{Audio-to-audio retrieval.}
For each generated RIR $\widehat{\mathbf{x}}_i$, we compute its normalized embedding 
$\widehat{\mathbf{A}}_i = \phi_A(\hat{\mathbf{x}}_i)$ and compare it to the normalized embeddings 
$\mathbf{A}_j = \phi_A(\mathbf{x}_j)$ of the ground-truth RIRs.
Similarity is measured as: %via cosine similarity:
\begin{equation}
    s_{ij} = \widehat{\mathbf{A}}_i^\top \mathbf{A}_j.
\end{equation}
The audio-to-audio recall at X (R@X) is then defined as:
\begin{equation}
\text{R@X} = \frac{1}{N} \sum_{i=1}^N 
\mathds{1} \left( i \in \operatorname{TopX}_j(s_{ij}) \right),
\end{equation}
where $\operatorname{TopX}_j(s_{ij})$ returns the indices of the X most similar ground-truth embeddings for each generated RIR. This metric measures how often the ground-truth RIR corresponding to a generated sample appears among the top-X most similar ground-truth embeddings.

\paragraph{Acoustic-to-geometry retrieval.}
We similarly evaluate the alignment between generated audio and its corresponding room geometry. 
For each generated RIR embedding $\phi_A(\hat{\mathbf{x}}_j)$, we compute cosine similarities with all ground-truth geometry embeddings $\phi_G(\mathbf{g}_i)$. 
The resulting recall at X (R@X) measures how often the correct geometry is retrieved among the top-X most similar embeddings.

\paragraph{Fréchet Distance in AGREE space.}
To assess distributional realism, we compute a Fréchet Distance (FD) between the generated and ground-truth RIR embeddings in the AGREE space. Let $\mathcal{N}(\mu_r, \Sigma_r)$ and $\mathcal{N}(\mu_g, \Sigma_g)$ denote multivariate Gaussian approximations of the ground-truth and generated audio embeddings, respectively:
\begin{equation}
    \begin{aligned}
    \mu_r &= \frac{1}{N_r}\sum_i \phi_A(\mathbf{x}_i), \\
    \Sigma_r &= \frac{1}{N_r - 1} \sum_i (\phi_A(\mathbf{x}_i) - \mu_r)(\phi_A(\mathbf{x}_i) - \mu_r)^\top.
    \end{aligned}
\end{equation}
and similarly for $\mu_g$, $\Sigma_g$. The Fréchet Distance is then defined as:
\begin{equation}
    \text{FD}_{G} = \|\mu_r - \mu_g\|_2^2 + \operatorname{Tr}\left(\Sigma_r + \Sigma_g - 2(\Sigma_r \Sigma_g)^{1/2}\right).
\end{equation}
Lower values indicate a closer match between the distributions of generated and real RIR embeddings, reflecting better geometry-consistent realism.

\begin{table}[t]
\centering
\caption{\textbf{Effect of the geometry and acoustic conditioning encoders (seen):} We evaluate different configurations of the geometry encoder $\phi_G$ and acoustic encoder $\phi_A$ on the seen set of the AcousticRooms dataset.
For $\phi_G$, we compare the ViT from xRIR and DINOv3 ViT-S/16 under various initialization strategies ($\mathcal{W}_{\text{DINO}}$): trained from scratch, frozen, or initialized with DINO weights. For $\phi_A$, we experiment with the ResNet-18 used in xRIR and our frozen VAE encoder.}

\resizebox{\linewidth}{!}{
    \begin{tabular}{cc|c|c|ccccc}
    \toprule
    \multicolumn{2}{c|}{\textbf{$\mathbf{\phi}_{\text{G}}$}} & \multirow{2}{*}{\textbf{$\mathbf{\phi}_{\text{A}}$}} & \multirow{2}{*}{\textbf{K}} & \multirow{2}{*}{\textbf{T60 (\%) $\downarrow$}} & \multirow{2}{*}{\textbf{C50 (dB) $\downarrow$}} & \multirow{2}{*}{\textbf{EDT (ms) $\downarrow$}} & \multirow{2}{*}{\textbf{R@5 (\%) $\uparrow$}} & \multirow{2}{*}{\textbf{FD$_G$ $\downarrow$}} \\
     
     \textbf{ViT} & \textbf{$\mathcal{W}_{\text{DINO}}$}  & & & & & & & \\
    \midrule

    \cite{liu2025haae} & \xmark & ResNet & 1 & 6.77 & 0.741 & 30.15 & 12.9 & 0.319 \\ 
    
    S/16 \cite{dinov3} & \xmark & ResNet & 1 & 6.83 & 0.740 & 30.23 & 17.65 & 0.310 \\ % 52500
    S/16 \cite{dinov3} & \SnowflakeChevron & ResNet & 1 & 7.61 & 0.956 & 36.98 & 6.40 & 0.365 \\ 
    \rowcolor{babyblue} S/16 \cite{dinov3} & \checkmark & ResNet & 1 & \textbf{6.46} & \textbf{0.692} & 28.11 & \textbf{21.54} & \textbf{0.296} \\ 
    
    S/16 \cite{dinov3} & \checkmark & VAE & 1 & 6.81 & \textbf{0.692} & \textbf{27.97} & 20.28 & 0.309 \\ 

    \midrule
    \cite{liu2025haae} & \xmark  & ResNet & 8 & 5.42 & 0.674 & 27.48 & 14.51 & 0.321 \\ 
    
    S/16 \cite{dinov3} & \xmark & ResNet & 8 & 5.60 & 0.685 & 27.51 & 19.24 & 0.311 \\ % 52500
    S/16 \cite{dinov3} & \SnowflakeChevron & ResNet & 8 & 6.30 & 0.858 & 32.42 & 7.62 & 0.374 \\ 
    \rowcolor{babyblue} S/16 \cite{dinov3} & \checkmark & ResNet & 8 & \textbf{5.32} & 0.643 & 25.69 & \textbf{21.89} & \textbf{0.299}   \\ 
    
    S/16 \cite{dinov3}  & \checkmark & VAE & 8 & 5.74 & \textbf{0.619} & \textbf{24.42} & 21.12 & 0.308 \\

    \bottomrule
    \end{tabular}
}

\label{tab: EncoderAblation_seen}
\end{table}
\begin{table}[t]
\centering
\caption{\textbf{Effect of DiT architecture variants (seen):} Performance comparison of different conditioning strategies on the seen set of the AcousticRooms dataset. We report results for In-Context, Cross-Attention (CA), and our hybrid design combining AdaLN for target information and CA for contextual cues.}

\resizebox{\linewidth}{!}{
    \begin{tabular}{l|c|ccccc}
    \toprule

      \textbf{Method} & \textbf{K} & \textbf{T60 (\%) $\downarrow$} & \textbf{C50 (dB) $\downarrow$} & \textbf{EDT (ms) $\downarrow$} & \textbf{R@5 (\%) $\uparrow$} & \textbf{FD$_G$ $\downarrow$}   \\

    \midrule
    In-Context & 1  & 72.37 & 11.299 & 1318.71 & 0.05 & 1.302 \\
    CA & 1 & 14.44 & 1.411 & 75.13 & 7.43 & 0.406 \\
    \rowcolor{babyblue} AdaLN+CA & 1 & \textbf{6.46} & \textbf{0.692} & \textbf{28.11} & \textbf{21.54} & \textbf{0.296} \\
    
    \midrule
    In-Context & 8 & \textbf{5.31} & 0.651 & 26.58 & 21.10 & 0.307 \\
    CA & 8 & 6.01 & 0.782 & 30.16 & 14.40 & 0.328 \\
    \rowcolor{babyblue} AdaLN+CA & 8 & 5.32 & \textbf{0.643} & \textbf{25.69} & \textbf{21.89} & \textbf{0.299} \\

    \bottomrule
    \end{tabular}
}

\label{tab: Architecture_seen}
\end{table}

\subsection{Baselines implementation details} \label{sub: baselinesdetails}
\paragraph{Random across rooms.} 
This baseline randomly selects one RIR from the entire dataset. In HAA, only training samples are considered.

\paragraph{Random same room.}
This baseline randomly selects one RIR from the same room. In HAA, only training samples from the same room are considered. 

\paragraph{Linear interpolation.}
This baseline interpolates the $K$ reference RIRs based on their distance to the target source. For each reference $k$, the distance $r_k$ to the target source $P_s^T$ is computed, and the weight is set as $w_k = 1 / (r_k + \epsilon)$, normalized to sum to 1. The final RIR is the weighted sum of the $K$ references. 

\begin{table*}[t]
\centering
\small
\caption{\textbf{Acoustic-to-geometry retrieval on the AcousticRooms dataset:} Results are shown for $K\in\{8,1,$\,\xmark$\}$ reference RIRs. For \methodname, we report mean and standard deviation over 5 generations. \abl~denotes ablations with either geometry (G) or audio conditioning removed. \methodname~achieves higher acoustic-to-geometry recall than the baselines, demonstrating higher scene consistency.}

\resizebox{\linewidth}{!}{

    \begin{tabular}{l|c|c|P{6pt}P{9pt}P{8pt}|P{6pt}P{9pt}P{8pt}} 
    \toprule
     \multirow{2}{*}{\textbf{Method}} & \multirow{2}{*}{\textbf{K} } &  \multirow{2}{*}{\textbf{G}} & \multicolumn{3}{c|}{\textbf{Unseen}} & \multicolumn{3}{c}{\textbf{Seen}} \\

     &  & &  \textbf{R@1 $\uparrow$} & \textbf{R@5 $\uparrow$} & \textbf{R@10 $\uparrow$} 
    & \textbf{R@1 $\uparrow$} & \textbf{R@5 $\uparrow$} & \textbf{R@10 $\uparrow$} \\
      
    \midrule
    Random Across Rooms & \xmark  & \xmark &
    0.03 & 0.13 & 0.21 &
    0.00 & 0.05 & 0.16 \\
    
    Random Same Room  & \xmark & \xmark &
    0.22 & 0.98 & 1.78 & 
    0.88 & 2.45 & 3.78 \\

    \methodname\abl & \xmark & \checkmark & 
    \textbf{5.12}\unc{0.15} & \textbf{16.10}\unc{0.19} & \textbf{24.01}\unc{0.12} &
    \textbf{5.58}\unc{0.10} & \textbf{16.18}\unc{0.12} & \textbf{23.09}\unc{0.20} \\ 

    \midrule 

    Nearest Neighbor & 1 & \xmark &
    0.05 & 2.04 & 4.09 &
    0.21 & 5.65 & 8.41 \\

    Fast-RIR & 1 & \checkmark &
    0.16 & 0.79 & 1.52 &
    0.15 & 0.82 & 1.61 \\

    xRIR & 1 & \checkmark &
    0.19 & 1.22 & 2.38 &
    0.16 & 1.27 & 2.54 \\

    \rowcolor{babyblue} \textbf{\methodname} & 1 & \checkmark & 
    \textbf{5.60}\unc{0.06} & \textbf{16.54}\unc{0.24} & \textbf{24.32}\unc{0.37} & 
    \textbf{6.37}\unc{0.08} & \textbf{17.71}\unc{0.21} & \textbf{25.05}\unc{0.10} \\ 

    \midrule
    
    Linear Interpolation & 8 & \xmark &
    0.28 & 1.40 & 2.87 & 
    0.64 & 2.85 & 5.10  \\
    
    Nearest Neighbor & 8 & \xmark & 
    0.16 & 8.66 & 15.20 &
    0.68 & \textbf{19.21} & \textbf{25.98}  \\

    \methodname\abl & 8 & \xmark &
    0.15\unc{0.01} & 0.45\unc{0.03} & 0.84\unc{0.07} &
    0.13\unc{0.01} & 0.43\unc{0.03} & 0.77\unc{0.03} \\

    Fast-RIR & 8 & \checkmark &
    0.17 & 1.09 & 1.82 &
    0.16 & 0.77 & 1.61 \\
    
    xRIR & 8 & \checkmark &
    0.43 & 1.85 & 3.01 &
    0.37 & 1.56 & 2.93 \\  % epoch 36

    \rowcolor{babyblue} \textbf{\methodname} & 8 & \checkmark &
     \textbf{5.79}\unc{0.07} & \textbf{16.84}\unc{0.08} & \textbf{24.62}\unc{0.17} &
     \textbf{6.53}\unc{0.14} & 18.10\unc{0.15} & 25.58\unc{0.15} \\
     
    \bottomrule
    \end{tabular}
}
\vspace{-8pt}

\label{tab: FS_AR_A2G}
\end{table*}

\paragraph{Nearest neighbor (KNN).}
From the $K$ reference RIRs, this baseline returns the RIR whose source position is closest (in Euclidean distance) to the target source position. 

\paragraph{Fast-RIR.}
This baseline uses the authors' original implementation. To align with our setup, we infer the room size from the panoramic depth map captured at the receiver pose and estimate $T_{60}$ using the $K$ contextual RIRs. Specifically, we compute $T_{60}$ for each of the $K$ RIRs and average the results. Following prior work~\cite{fewshot}, we incorporate an energy decay loss to improve performance.

\paragraph{xRIR.}
We use the implementation provided by the authors. Following their supplementary material, the Vision Transformer encoder uses 6 multi-head attention layers (8 heads, hidden size 512) with a patch size of $16\times32$. 
Poses are encoded using sinusoidal positional embeddings applied to each 3D coordinate with 20 frequency bins. We set $\lambda = 0.01$ to balance the STFT and energy-decay losses.  
We train xRIR on the AcousticRooms training split, excluding both the seen and unseen test sets, and use the same trained model for evaluating both splits.
For HAA, we fine-tune the AcousticRooms-pretrained model on the four HAA rooms at the same time.

\paragraph{INRAS.}
Similar to~\cite{RAF}, we modify the original bounce point sampling strategy, which only sampled points at a fixed height. Instead, we construct scene meshes from the HAA surface annotations and apply Poisson sampling to obtain 256 3D bounce points, providing a richer representation of the scene geometry. As the training set contains only 12 RIRs per room, we train per scene with a batch size of 12 for 5k epochs. We found that adjusting the multi-resolution STFT loss parameters improves performance, using FFT sizes $\{128, 512, 1024, 2048\}$, hop lengths $\{16, 50, 120, 240\}$, and window lengths $\{80, 240, 600, 1200\}$. %The rotation is fixed to $0^\circ$ to match the HAA setup.

\paragraph{DiffRIR.}
We use the official implementation and adapt it to 22{,}050\,Hz. Specifically, we reduce the maximum predicted audio length from 96{,}000 samples at 48\,kHz to 44{,}100 samples at 22.05\,kHz. We use the authors’ precomputed sound trajectories and rescale the delay values to match the new sample rate.

\section{Additional results} \label{sec: SM_Results}
\subsection{Seen set of the AcousticRooms dataset} \label{sub: seenset}
\paragraph{Comparison to the baselines.}
\cref{tab: FS_AR_seen} reports results on the seen set of the AcousticRooms dataset, which contains novel source-receiver positions within scenes observed during training. With $K{=}8$, \methodname~reduces errors by $23.9\%$, $29.8\%$, and $24.8\%$ on T60, C50, and EDT, respectively, compared to xRIR. It also substantially improves scene-consistency metrics and slightly improves FD$_G$ ($-8.8\%$) over xRIR. 
Consistent with the unseen-set results, \methodname~with $K{=}1$ outperforms all other one-shot methods and even surpasses xRIR with $K{=}8$.

\paragraph{Ablation of conditioning modalities.}
\cref{tab: FS_AR_seen} reports \methodname~variants with one conditioning modality removed (\abl). The trends mirror those observed on the unseen set. Removing geometry leads to large drops in geometry-related metrics (R@1-10, FD$_G$) even with $K{=}8$ contextual RIRs. Conversely, using geometry without contextual RIRs still leads to satisfactory performance. C50 and EDT are better with geometry-only than with context-only conditioning, consistent with their dependence on early reflections, which are closely tied to local scene geometry. T60 is better with audio-only conditioning, aligning with its dependence on global room characteristics that are difficult to infer from local geometry alone.

\paragraph{Ablation on geometry and acoustic encoders.}
\cref{tab: EncoderAblation_seen} compares \methodname~with different geometry and acoustic encoders on AcousticRooms'seen set. Consistent with the unseen-set results, fine-tuning the DINOv3 ViT S/16 achieves the best overall performance. Using our frozen VAE as the acoustic encoder improves C50 and EDT but slightly degrades other metrics, further motivating the use of the simpler and more efficient ResNet-18 for acoustic conditioning.

\paragraph{Influence of the DiT architecture.}
Consistent with results on the unseen set of the AcousticRooms dataset, the combination of AdaLN for target-related conditioning and cross-attention for context leads to the best performance (see \cref{tab: Architecture_seen}).

\subsection{Reversed setup in HAA}
The HAA setup is reversed compared to AR. In HAA, context references share the same source, whereas in AR they share the same receiver. Consequently, the panoramic depth is captured at the source pose in HAA and at the receiver pose in AR. In the main paper, we fine-tune FLAC on HAA by simply swapping source and receiver poses (same for AGREE).
We also evaluated inverting the receiver projection. This modification significantly improves T60/C50/EDT metrics but reduces scene consistency metrics. In this setting, AGREE used for evaluation is fine-tuned with the same modification. Results are reported in \cref{tab: HAAReverseSetup}

\begin{table}[t]
\centering
\caption{\textbf{\amandine{Sim-to-real transfer on the Hearing-Anything-Anywhere dataset with inverted receiver pose:}} Few-shot methods are compared to Diff-RIR and INRAS, which require per-scene training ($^\dag$). For \methodname, we report the mean and standard deviation over 5 generations. Inverting the receiver pose improves T60/C50/EDT but reduces geometry-consistency metrics.}

\resizebox{\linewidth}{!}{

    \newcolumntype{P}[1]{>{\hspace{0pt}}c<{\hspace{#1}}}
    \begin{tabular}{l|c|P{2pt}P{5pt}P{2pt}cc}
    \toprule
    \textbf{Method} & \textbf{K} & \textbf{T60 (\%)} $\downarrow$ & \textbf{C50 (dB)} $\downarrow$ & \textbf{EDT (ms)} $\downarrow$ & \textbf{R@5 (\%)} $\uparrow$ & \textbf{FD}$_G$ $\downarrow$ \\
    
    \midrule
    Random Across Rooms & \xmark & 17.40 & 10.283 & 533.99 & 2.02 & 0.418 \\

    Random Same Room & \xmark & 8.00 & 4.805 & 180.15 & 1.51 & 0.253 \\

    \midrule
    Nearest Neighbor & 1 & 8.19 & 5.000 & 187.55 & 1.32 & \textbf{0.268} \\
    xRIR & 1 & 8.63 & 4.862 & 183.27 & 8.43 & 0.372 \\

    \rowcolor{babyblue} \textbf{\methodname} & 1 & \textbf{3.02}\unc{0.04} & \textbf{1.676}\unc{0.023} & \textbf{75.14}\unc{0.98} & \textbf{10.31}\unc{1.10} & 0.899 \\

    \midrule
    Linear Interpolation & 8 & 4.12 & 2.695 & 88.19 & 8.24 & 0.824 \\
    Nearest Neighbor & 8 & 2.89 & 1.923 & 77.24 & \textbf{12.64} & \textbf{0.243} \\
    xRIR & 8 & 6.53 & 3.492 & 149.69 & 10.93 & 0.383  \\ % paper setup
    \rowcolor{babyblue} \textbf{\methodname}  & 8 & \textbf{2.87}\unc{0.04} & \textbf{1.595}\unc{0.017} & \textbf{69.14}\unc{0.56} & 10.31\unc{0.75} & 0.936 \\%\unc{0.001} \\ 

    \midrule
    INRAS$^{\dag}$ & 12 & 6.61 & 3.967 & 158.07 & 2.79 & 1.070 \\
    Diff-RIR$^{\dag}$ & 12 & 3.74 & 2.067 & 88.09 & 18.03 & 0.539 \\ % max len 44100

    \bottomrule
    \end{tabular}
}

\label{tab: HAAReverseSetup}
\end{table}

\subsection{Additional metrics}
\paragraph{Acoustic-to-geometry retrieval.}
To complement the audio-to-audio retrieval metrics presented in the main document, we provide acoustic-to-geometry retrieval metrics in \cref{tab: FS_AR_A2G} for both the seen and unseen sets of the AcousticRooms dataset.
\methodname~produces more geometry-consistent RIRs as demonstrated by its superior recall performances. 

\paragraph{MAG and ENV on HAA.}
\begin{table}[t]
\centering
\small 

\caption{\textbf{Additional metrics on the Hearing-Anything-Anywhere dataset: } Few-shot methods are compared against Diff-RIR, which requires per-scene training ($^\dag$) using MAG and ENV error.}

\resizebox{0.6\linewidth}{!}{
    \begin{tabular}{l|c|cc}
    \toprule
    \textbf{Method} & \textbf{K} & \textbf{MAG} $\downarrow$ & \textbf{ENV} $\downarrow$  \\
    
    \midrule
    Random Across Rooms & \xmark & 6.97 & 1.581  \\
    Random Same Room & \xmark & 3.22 & 0.470 \\

    \midrule
    Nearest Neighbor & 1 & 3.24 & 0.475 \\
    xRIR & 1 & 3.97 & 0.577 \\
    \rowcolor{babyblue} \textbf{\methodname} & 1 & \textbf{3.16} & \textbf{0.393}  \\

    \midrule
    Linear Interpolation & 8 & 3.63 & 0.835 \\
    Nearest Neighbor & 8 & \textbf{2.88} & \textbf{0.373} \\
    xRIR & 8 & 3.44 & 0.407 \\
    
    \rowcolor{babyblue} \textbf{\methodname}  & 8 & 3.11 & 0.381 \\

    \midrule
    INRAS$^{\dag}$ & 12 & 3.30 & 0.654  \\
    Diff-RIR$^{\dag}$ & 12 & 2.53 & 0.352  \\ % max len 44100

    \bottomrule
    \end{tabular}
}
\vspace{-8pt}

\label{tab: FS_HAA_MAGENV}
\end{table}

We report MAG and ENV metrics on the HAA dataset in \cref{tab: FS_HAA_MAGENV}.
Following \cite{HAA}, MAG denotes the multiscale log-spectral L1 distance, which compares generated and ground-truth waveforms in the time-frequency domain across multiple resolutions. ENV is the envelope distance, defined as the L1 distance between the log-energy envelopes of the generated and ground-truth waveforms. The energy decay envelope captures the decay characteristics of a RIR, reflecting the reverberant properties of a room. For both metrics, we follow diffRIR implementation. Consistent with other metrics, \methodname~outperforms xRIR at $K{=}8$, and surpasses others at $K{=}1$.

\subsection{Timestep sampling strategy} \label{subsub: timestepsampler}
In \cref{tab: FS_AR_Timestep}, we compare the effect of different training timestep sampling strategies on performance: (i) \textit{LogSNR}, our baseline, which emphasizes higher $t$ values; (ii) \textit{Uniform}, sampling $t \sim \mathcal{U}(0,1)$; (iii) \textit{Ones}, fixing $t=1$ (fully noisy); and (iv) \textit{Logit-Normal}, where $t = \sigma(\alpha)$ with $\alpha \sim \mathcal{N}(0,1)$, concentrating $t$ in the mid-range.

\begin{table}[h!]
\centering
\caption{\textbf{Comparison of timestep sampling strategies during training on the generation performance:} \textit{LogSNR} emphasizes high $t$ values, \textit{Logit-Normal} concentrates on intermediate values, \textit{Ones} fixes $t=1$ (\ie, full noise), and \textit{Uniform} samples $t$ uniformly. \textit{LogSNR} provides the best overall trade-off between seen and unseen performance.}

\resizebox{\linewidth}{!}{
    \begin{tabular}{l|ccc|ccc}
    \toprule
     \multirow{2}{*}{\textbf{Timestep Sampler}} & \multicolumn{3}{c|}{\textbf{Unseen}} & \multicolumn{3}{c}{\textbf{Seen}} \\

     & \textbf{T60 (\%) $\downarrow$} & \textbf{C50 (dB) $\downarrow$} & \textbf{EDT (ms) $\downarrow$} & \textbf{T60 (\%) $\downarrow$} & \textbf{C50 (dB) $\downarrow$} & \textbf{EDT (ms) $\downarrow$} \\
     
    \midrule
    
    Uniform & 8.97 & \textbf{0.943} & \underline{38.18} & \textbf{5.22} & \textbf{0.636} & 25.83 \\
    
    Ones & \textbf{8.17} & 1.044 & \underline{37.48} & \underline{5.29} & \textbf{0.636} & \textbf{25.60} \\
    
    Logit-Normal & 9.77 & 1.068 & 42.65 & 5.80 & 0.681 & 27.98 \\

     \rowcolor{babyblue} LogSNR & \underline{8.60} & \underline{0.970} & \textbf{37.13} & 5.32 & \underline{0.643} & \underline{25.69} \\

    \bottomrule
    \end{tabular}
}		
					
\label{tab: FS_AR_Timestep}
\end{table}

\begin{figure*}[t]
    \centering
    \includegraphics[width=0.88\linewidth]{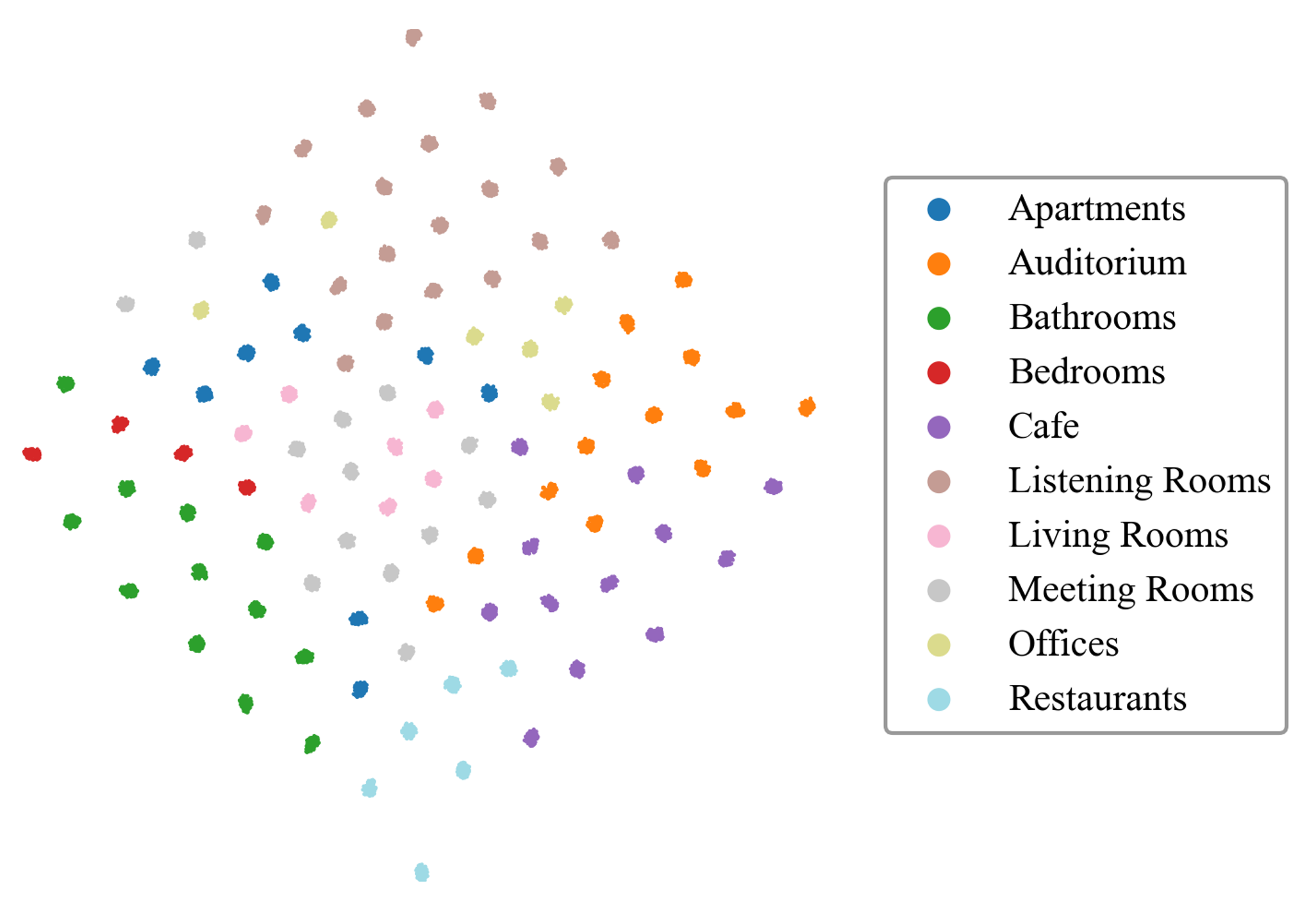}
    \caption{\textbf{t-SNE visualization of generated RIRs across unseen rooms:} Each color represents a different room category. Samples cluster tightly within the same conditioning and remains well separated across conditionings and rooms, confirming consistent yet diverse RIR generation.}
    \label{fig: TSNE}
\end{figure*}

\section{Qualitative results} \label{sec: SM_quali}
\paragraph{t-SNE.}
\cref{fig: TSNE} provides a t-SNE visualization of multiple generated samples across unseen rooms. Samples with the same conditioning cluster tightly, while those from different rooms or conditionings are clearly separated, further demonstrating that the model captures both consistency and diversity in its generations. 
We observe that acoustically similar scenes lies next to the other, \eg, Restaurants-Cafes, Auditoriums-Listening Rooms, and Apartments (which include bathrooms)-standalone Bathrooms.

\begin{figure*}
    \centering
     \includegraphics[width=\linewidth]{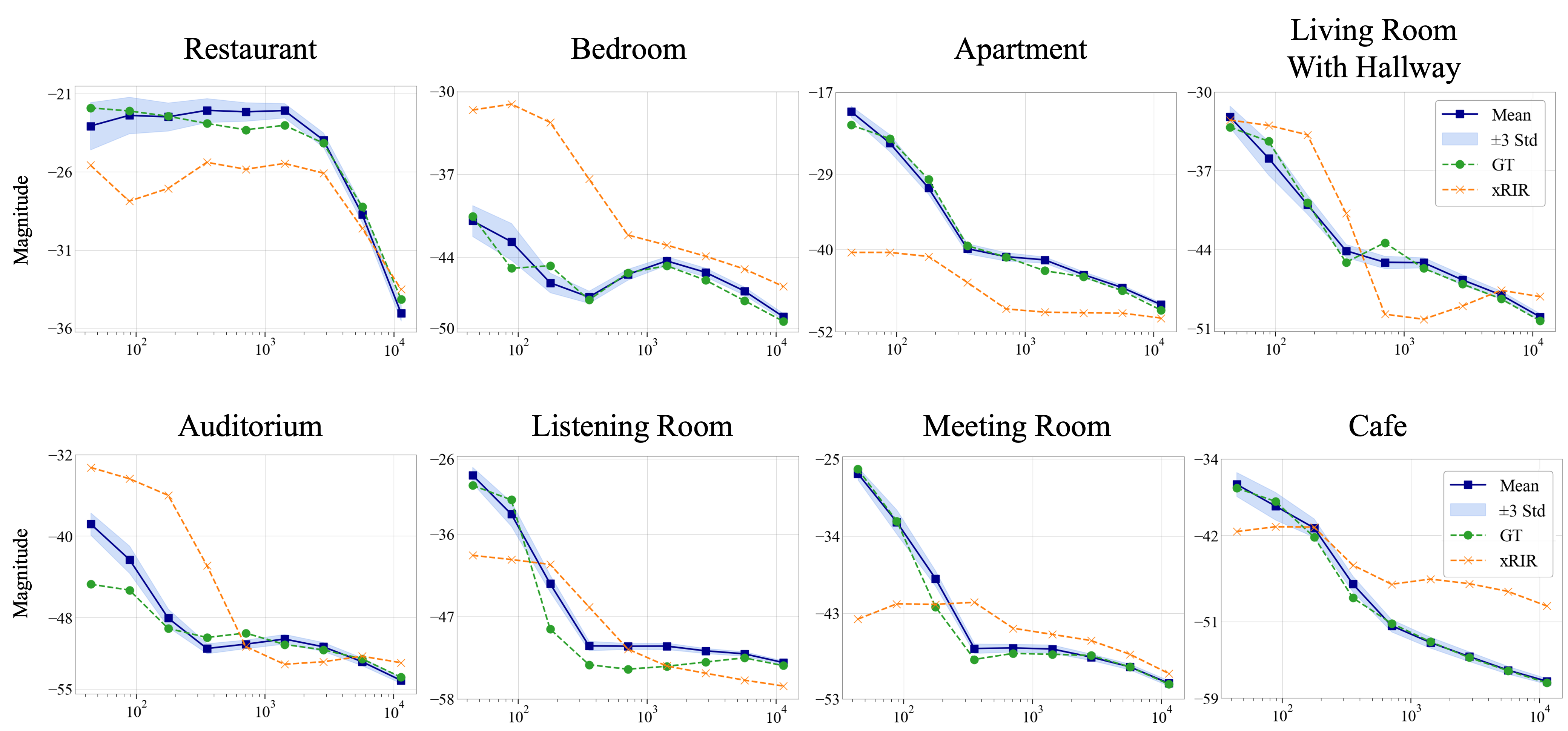}
    \caption{\textbf{Additional octave band analysis on the AcousticRooms dataset:} We generate 100 samples per instance in the unseen set with \methodname~under identical conditioning, and plot the mean along with a $\pm3$ standard deviation interval (covering 99.7\% of the distribution). Ground truth and xRIR predictions are shown for comparison. }
    \label{fig:MoreOctaveBandP1}
\end{figure*}

\begin{figure*}
    \centering
    \includegraphics[width=\linewidth]{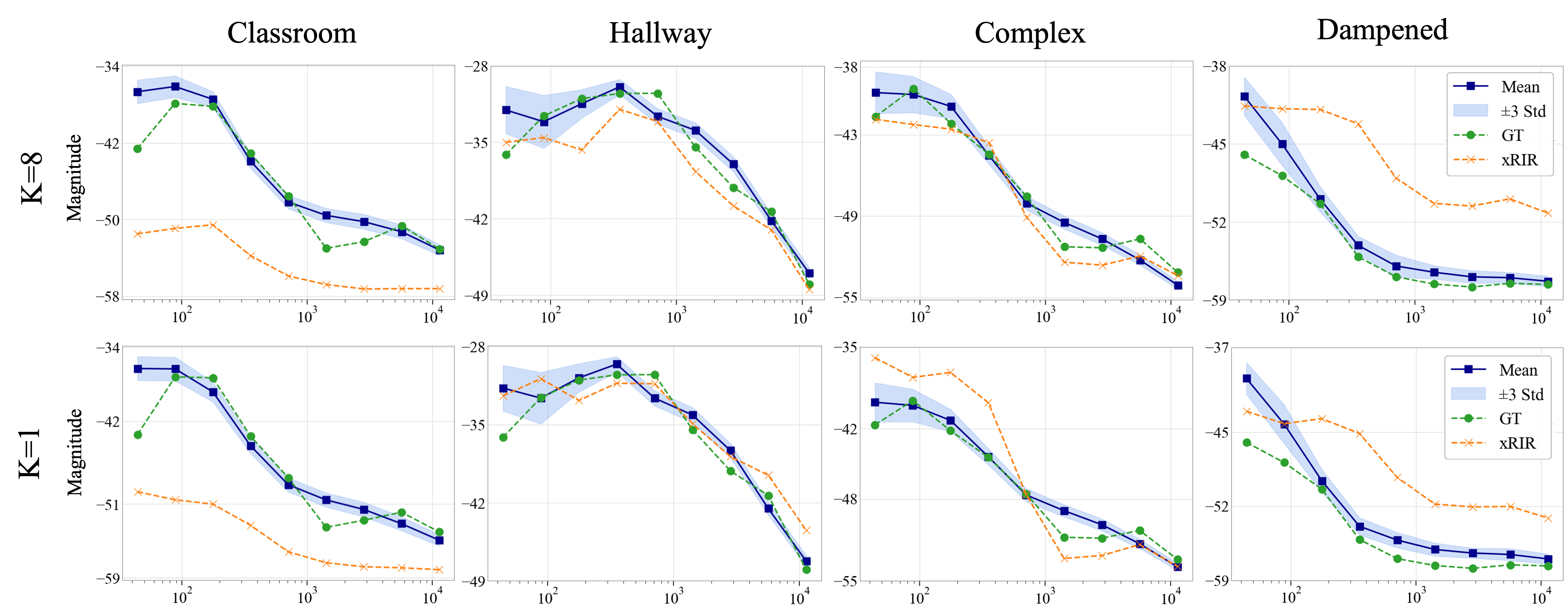}
    \caption{\textbf{Octave band analysis on the HAA dataset.}}
    \label{fig:HAAOctaveBand}
\end{figure*}

\paragraph{Octave-band analysis.}
We provide more visualization of octave-band analysis in \cref{fig:MoreOctaveBandP1},  \cref{fig:HAAOctaveBand}.

\begin{figure*}
    \centering
    \includegraphics[width=\linewidth]{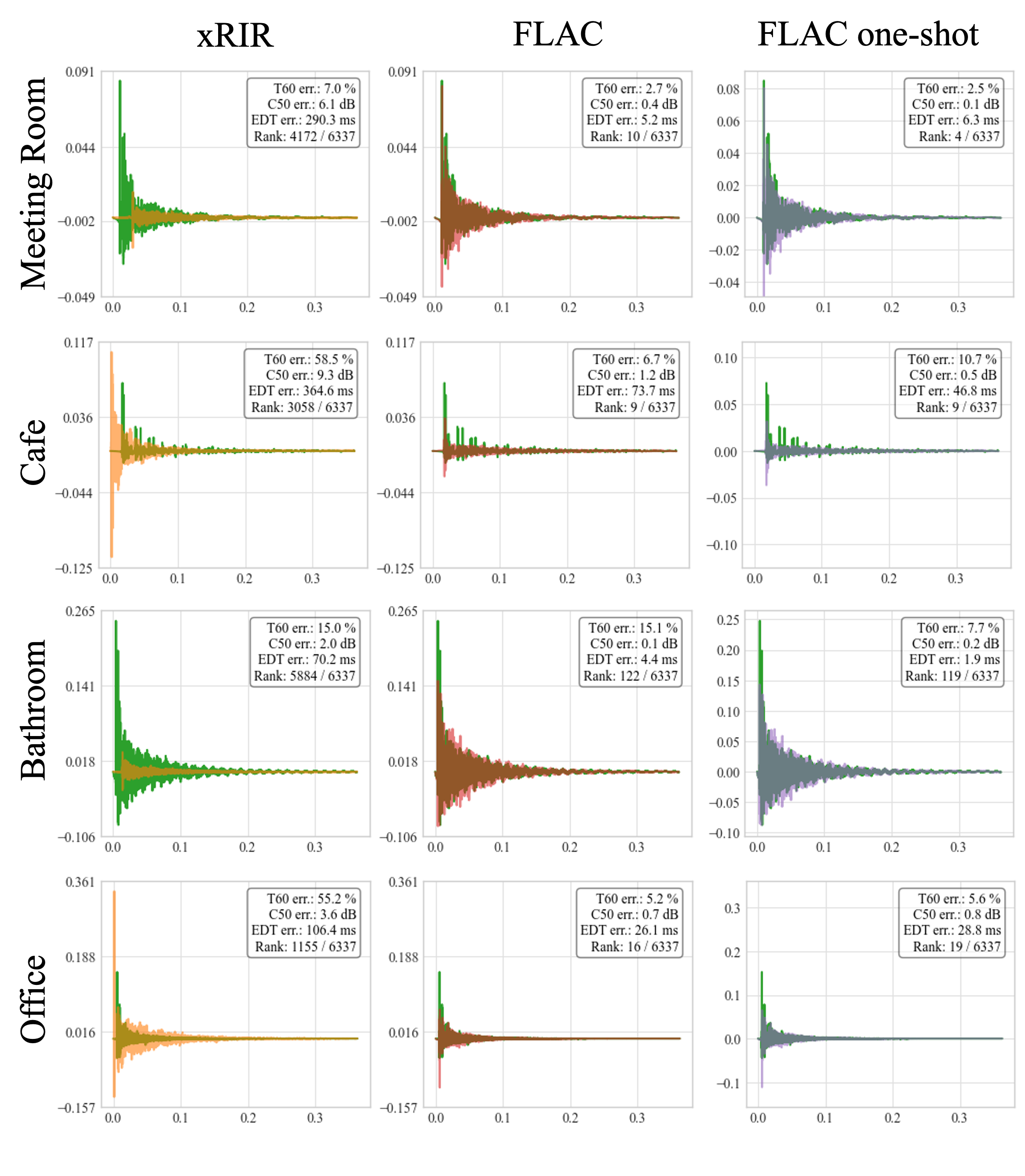}
    \caption{\textbf{Qualitative comparison of predicted RIR waveforms on the unseen set of the AcousticRooms dataset:} We compare xRIR (orange), \methodname~(red), and \methodname~with $K=1$ (purple) against the ground truth (green).}
    \label{fig:qualiwaveformsAR}
\end{figure*}

\begin{figure*}
    \centering
    \includegraphics[width=\linewidth]{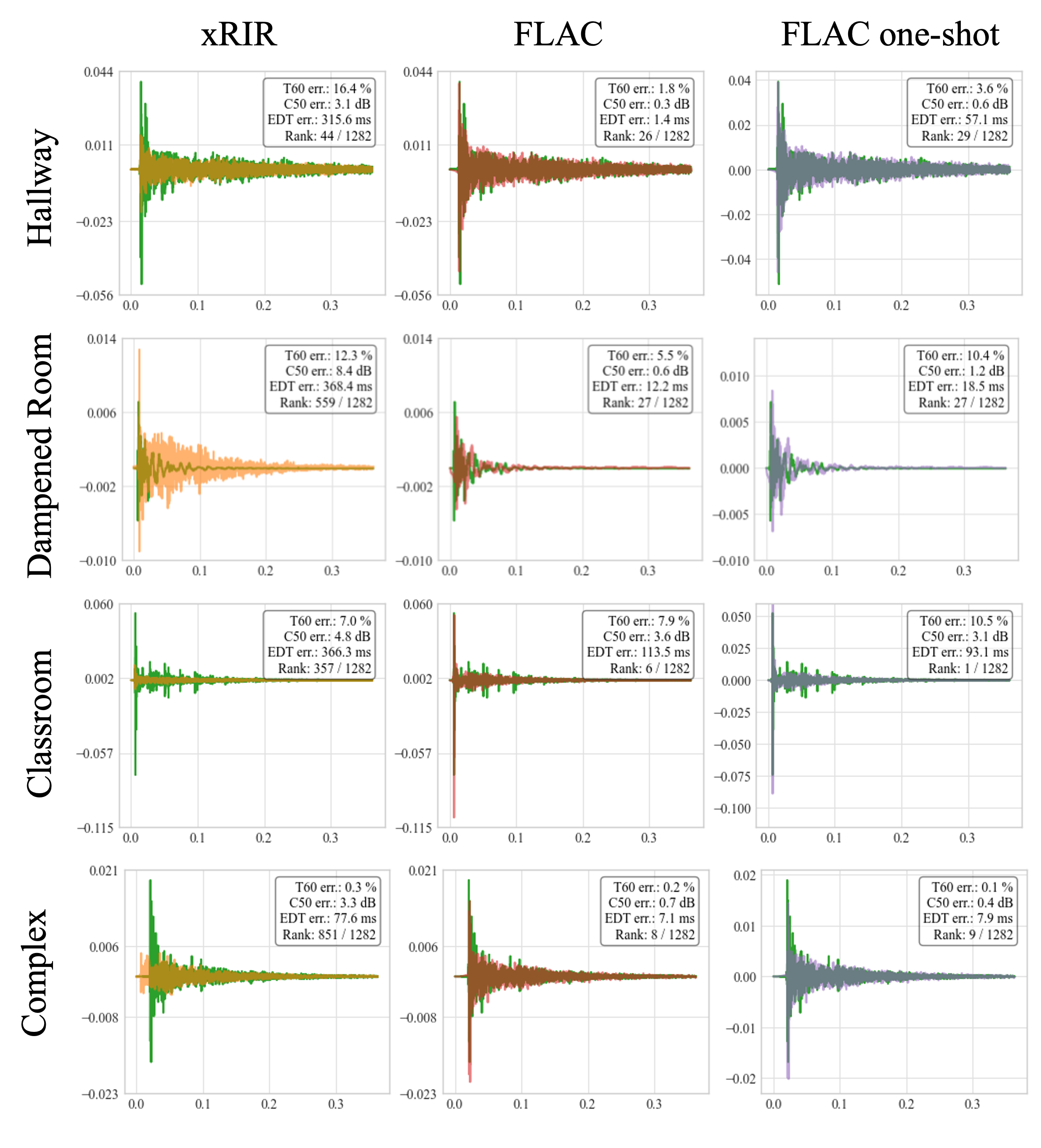}
    \caption{\textbf{Qualitative comparison of predicted RIR waveforms on the HAA dataset:} We compare xRIR (orange), \methodname~(red), and \methodname~with $K=1$ (purple) against the ground truth (green).}
    \label{fig:qualiwaveformsHAA}
\end{figure*}

\paragraph{Waveforms.}
We present in \cref{fig:qualiwaveformsAR} and \cref{fig:qualiwaveformsHAA} generated waveforms with xRIR and \methodname~against the ground truth for different scenes of the AcousticRooms and HAA datasets. 
The rank metric corresponds to the audio-to-audio retrieval rank of each prediction against all the ground truth RIRs of the evaluation set.

\paragraph{Video.}
We provide in the project page several examples of audio generated with \methodname~using $K{=}8$ and $K{=}1$ reference RIRs in unseen scenes from both simulated and real environments.  
The anechoic speech samples used for auralization come from the EARS dataset \cite{richter2024ears}, and the music samples are taken from AVAD-VR \cite{AnechoicAudio}.  
For audio rendering along trajectories in real scenes, we convolve our predicted single-channel RIR at each point of the trajectory with a head-related impulse response derived from a predefined head-related transfer function. This produces binaural RIRs, which are then convolved with the source audio to obtain the final binaural rendering.  
The rank metric shown in the video corresponds to the audio-to-audio retrieval rank of each prediction against all ground-truth RIRs in the full unseen set, \ie, the position of the correct ground-truth RIR in the similarity matrix.

\section{Perceptual evaluation} \label{sec: SM_perceptual_eval}
\begin{figure}[h]
    \centering
        \includegraphics[width=\linewidth]{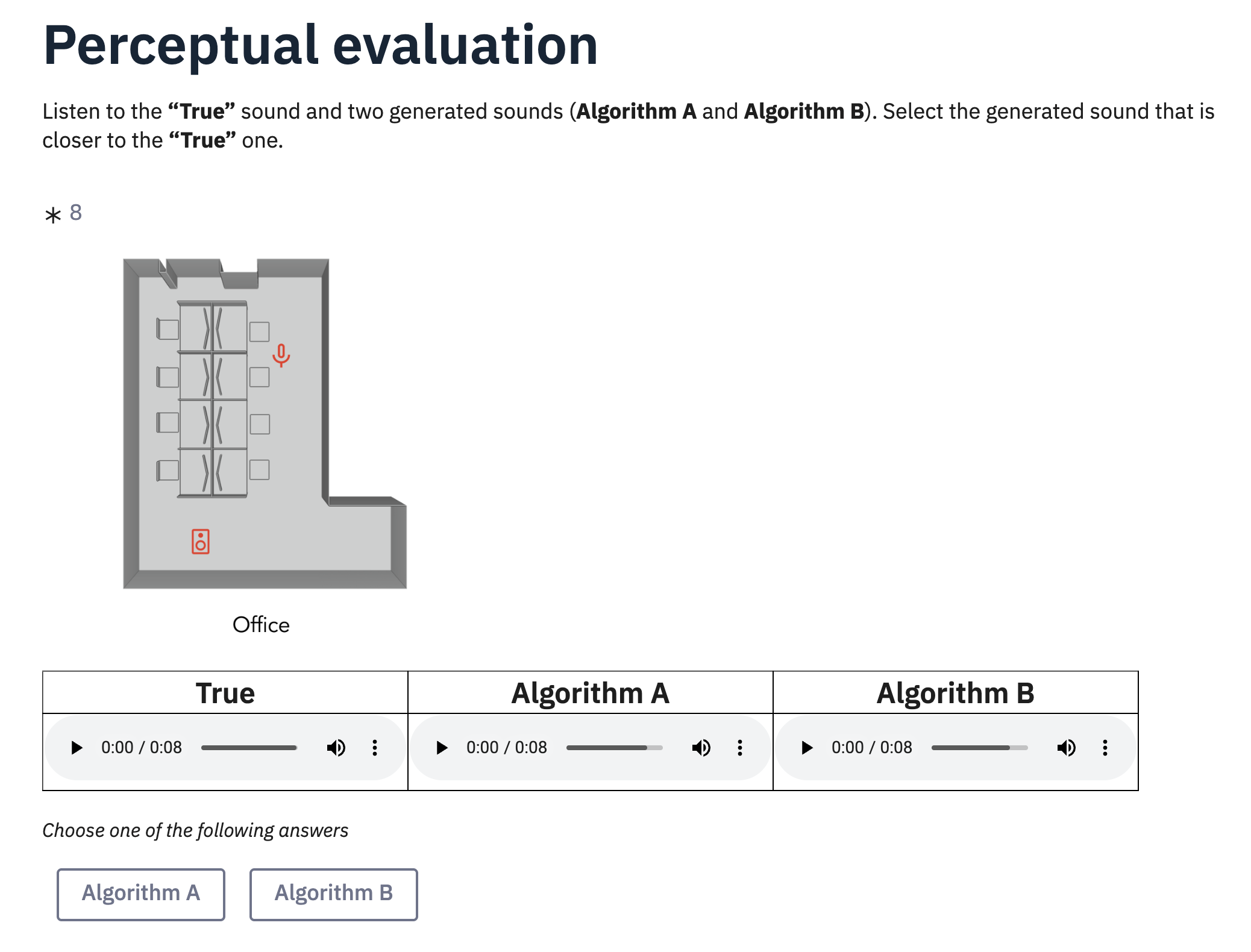}
        \caption{\textbf{User study interface.} We created 2 synthetic audio using FLAC one-shot and xRIR one-shot and ask people which one matches closely with the ground truth.}
        \label{fig:interface}

\end{figure}

We conducted a listening study with 46 participants on 14 unseen AR scenes. Participants were presented with the ground-truth (GT), audio generated by FLAC (1-shot) and xRIR (8-shot), and were asked to select which audio sounded closer to the GT. FLAC was preferred in 93.01\% of cases. The order of questions and the assignment of methods to “algorithm A” and “algorithm B” were randomized. Participants were first shown a top view of the scene including the microphone and source positions. They then listened to the GT audio ("true"), obtained by convolving the ground-truth RIR with an anechoic signal. Next, they listened to the two generated samples (“algorithm A” and “algorithm B”), which were produced by convolving the same anechoic signal with RIRs generated by FLAC or xRIR. \cref{fig:interface} shows an example of the user interface used.

For the audio content, we used anechoic speech from the EARS dataset \cite{richter2024ears}. Different voices were used across questions, and some questions included music excerpts from AVAD-VR \cite{AnechoicAudio} to evaluate additional use cases. To ensure reliable listening conditions, participants were required to complete the survey on a laptop or desktop computer while wearing headphones. They were also asked to report their background noise environment: 34 participants reported a quiet environment and 12 reported some background noise (possibilities were "quiet", "some background noise", "very noisy").

A total of 49 participants initially completed the survey. To screen for potential hearing issues or non-compliance with headphone use, we asked participants whether they had any known hearing impairments and whether they were wearing headphones. Additionally, the first question was a control trial in which participants had to choose the closer audio to the GT audio between the same GT audio and an audio recorded in a completely different scene. Three participants failed this control question and were excluded from the analysis, leaving 46 valid participants. Errors on this control question were correlated with participants reporting hearing impairments.

We also collected demographic information. Among the 46 participants, 13 identified as female and 33 as male. The age distribution was: 3 under 18, 6 between 18-24, 17 between 25-34, 5 between 35-44, 9 between 45-54, 4 between 55-64, and 2 aged 65 or older.

\newpage
\section{Number of parameters and inference speed} \label{sec: SM_param}
We report in \cref{tab: FS_ParamSpeed} the trainable and total inference parameter counts of our models, xRIR, Fast-RIR, and INRAS, along with inference times measured on a single RTX 4090 GPU. Note that INRAS must be trained for each scene, this results in 240M parameters for the full Acoustic Rooms dataset. Results are shown for variants with the frozen VAE encoder (replacing the jointly trained ResNet-18), with xRIR's ViT architecture instead of DINOV3 ViT-S/16, with a DiT depth of 4 instead of 12 (FLAC-S), and for the VAE used to obtain the latent $\mathbf{z}_0$. Our models achieve real-time performance, as inference is faster than the duration of the generated audio. 
We also compare the performance of FLAC-S against FLAC and xRIR in \cref{tab: FLAC-s}. FLAC-S (39M), obtained by reducing the DiT depth from 12 to 4, has a similar number of parameters to xRIR (32M). Despite this reduction, FLAC-S maintains performance comparable to FLAC, outperforming xRIR. This indicates that the gains of our method are not driven by model size.
%\newpage

\begin{table}[h]
\centering
\caption{\textbf{Number of trainable parameters along with inference parameters and speed} for our model, its variants, and state-of-the-art methods}. M denotes million.
\resizebox{\columnwidth}{!}{
\begin{tabular}{l|ccc}
\toprule
      Model & Trainable Param. (M) & Inference Param. (M) & Speed (ms) \\ 
      \midrule
       INRAS \cite{INRAS} & 1$\times N_{\text{scene}}$ & 1$\times N_{\text{scene}}$ & 1.9 \\
      Fast-RIR \cite{FastRIR} & 116 & 116 & 0.6  \\
      xRIR \cite{liu2025haae} & 32.1 & 32.1 & 38.5 \\
      \midrule
      VAE & 14.6 & 14.2 & 8.9  \\ % time to encode and decode
      \methodname~VAE & 38.6 & 45.7 & 13.5 \\
      \methodname~ViT \cite{liu2025haae} & 48.5 & 55.6 & 13.7 \\
      \methodname-S & 37.2 & 44.3 & 7.0  \\
      \rowcolor{babyblue} \methodname & 50.3 & 57.4 & 13.5 \\
     
\bottomrule
\end{tabular}
}
\label{tab: FS_ParamSpeed}
\end{table}

\begin{table}[h]
\centering
\caption{\textbf{Comparison between FLAC-S, FLAC, and xRIR on the unseen AcousticRooms set}: FLAC-S reduces the parameter count by 13M compared to FLAC. Despite having a similar number of parameters to xRIR, it achieves performance comparable to FLAC, outperforming xRIR.}
\newcolumntype{P}[1]{>{\hspace{0pt}}c<{\hspace{#1}}}
\resizebox{\linewidth}{!}{
    \begin{tabular}{l|c|P{2pt}P{5pt}P{2pt}cc} %c}
    \toprule
    
     \textbf{Method} & \textbf{K} & \textbf{T60 (\%) $\downarrow$} & \textbf{C50 (dB) $\downarrow$} & \textbf{EDT (ms) $\downarrow$} & \textbf{R@5 (\%) $\uparrow$} & \textbf{$\text{FD}_G$ $\downarrow$} \\

    \midrule 
    xRIR & 1 &
    14.47 & 1.961 & 74.45 & 1.36 & 0.263 \\ 
    
    FLAC-S & 1 &
    \textbf{9.71}\unc{0.04} & \textbf{1.010}\unc{0.002}  & \textbf{39.58}\unc{0.18} & \textbf{17.80}\unc{0.11} & \textbf{0.264} \\ 
    
     FLAC & 1 & 
     9.95\unc{0.05} & 1.046\unc{0.002} & 40.04\unc{0.22} & 18.92\unc{0.10} & 0.303 \\ 

    \midrule
    xRIR & 8  &
    9.98 & 1.354 & 49.40 & 2.00 & 0.307 \\
    
    FLAC-S & 8 &
    8.69\unc{0.02} & \textbf{0.942}\unc{0.001}  & 37.14\unc{0.08} & 17.88\unc{0.17} & \textbf{0.267} \\ 

    FLAC & 8 &
    \textbf{8.60}\unc{0.01} & 0.970\unc{0.002}  & \textbf{37.13}\unc{0.02} & \textbf{19.38}\unc{0.15} & 0.305 \\ 
   
    \bottomrule
    \end{tabular}
}

\label{tab: FLAC-s}
\end{table}

\clearpage
\clearpage

\begin{comment}
\newpage
\twocolumn[
\begin{center}
\addcontentsline{toc}{section}{References}
\end{center}
]

{
    \small
    \bibliographystyle{ieeenat_fullname}
    \bibliography{main}
}
\end{comment}

\end{document}